\documentclass[a4paper,11pt]{article}
\pdfoutput=1 

\usepackage{jheppub} 
\usepackage{siunitx}                     
\usepackage{cleveref}                     

\usepackage[T1]{fontenc}
\usepackage{import}
\usepackage{physics}
\usepackage{siunitx}
\usepackage{graphics}
\usepackage{caption}
\usepackage{subcaption}
\usepackage{collcell}
\usepackage{placeins}

\title{Di-electron production at the LHC: Unravelling virtual-photon and heavy-flavour contributions}

\author[a]{Anton Andronic}
\author[b]{\!\!, Tom\'a\v{s} Je\v{z}o}
\author[b]{\!\!, Michael Klasen}
\author[a]{\!\!, Christian Klein-Bösing}
\author[b]{and Alexander Puck Neuwirth}

\affiliation[a]{Institut  für  Kernphysik,  Universität Münster,  Wilhelm-Klemm-Straße 9, 48149 Münster, Germany}
\affiliation[b]{Institut  für  Theoretische  Physik, Universität Münster,  Wilhelm-Klemm-Straße 9, 48149 Münster, Germany}

\emailAdd{andronic@uni-muenster.de}
\emailAdd{tomas.jezo@uni-muenster.de}
\emailAdd{michael.klasen@uni-muenster.de}
\emailAdd{christian.klein-boesing@uni-muenster.de}
\emailAdd{alexander.neuwirth@uni-muenster.de}

\abstract{The production of virtual photons is a very sensitive probe of the properties of the quark-gluon plasma. As they are experimentally detected by lepton pairs, they suffer from a large background arising from hadron decays. Light-flavour hadrons dominate at low invariant masses below $m_{ee}\sim0.5$ GeV and heavy flavours above. These contributions must therefore also be taken into account in experimental analyses at the LHC. In this paper, we calculate the direct contribution from virtual photons produced in the Drell-Yan process with an additional jet in POWHEG and find that it is significant at low invariant masses. We also simulate the background contributions from $c\bar c$ and $b \bar b$ production with POWHEG and quantify the theoretical uncertainties due to variations of the perturbative scales and parton distribution functions. We find larger relative and absolute uncertainties for the lighter $c$ quarks than for heavier $b$ quarks.}

\preprint{MS-TP-23-22}

\keywords{Perturbative QCD, hadron colliders, lepton pairs, Drell-Yan, heavy flavours}

\begin{document} 
\maketitle
\flushbottom

\section{Introduction}
\label{sec:intro}

One of the key objectives of the Large Hadron Collider (LHC) physics program is the understanding of high-energy density and temperature conditions in ultra-relativistic heavy-ion collisions. These collisions lead to the creation of a quark-gluon plasma (QGP), a system of quarks and gluons in a deconfined state with restored chiral symmetry\,\cite{Rapp:2009yu} that is believed to have existed in the early Universe shortly after the Big Bang and which is predicted by quantum-chromodynamic (QCD) calculations on the lattice\,\cite{Karsch:1998qj}. In the high-temperature and low baryochemical potential region of the QCD matter phase diagram, this state of matter is reached in a smooth crossover, while at low temperatures and high baryochemical potentials it is expected to be reached in a first-order phase transition with a discontinuity in the thermodynamic variables \cite{Philipsen:2019rjq}. 
One of the expected signals of a QGP is the radiation of ``thermal photons'', with a particular transverse momentum spectrum reflecting its temperature\,\cite{Rapp:2009yu,Tserruya:2009zt}.

Photons are widely regarded as indispensable probes of heavy-ion collisions. Generated throughout all stages of the collision process, they originate either from {\em decays} of light- and heavy-flavour hadrons or are produced {\em directly} from hard-scattering processes like the quark-gluon Compton scattering and quark--anti-quark annihilation. In addition to these sources, the hot medium produced in heavy-ion collisions may emit thermal radiation, commonly grouped together with the {\em direct} production. Unaffected by the strong interaction, the photons traverse the dense medium without undergoing any final-state interactions. In contrast to real photons, virtual photons ($\gamma^* \to e^+e^-$) carry a mass ($m_{ee}$), unaltered by the radial flow, that can act as an approximate clock, allowing for the separation of different collision stages. The pre-equilibrium effects become evident around $m_{ee} \gtrsim \SI{2}{\giga\electronvolt}$, followed by the expected formation of the QGP in the range $\SI{1}{\giga\electronvolt} \lesssim m_{ee} \lesssim \SI{2}{\giga\electronvolt}$, and finally the hadronic phase at later emission stages with $ m_{ee} \lesssim \SI{1}{\giga\electronvolt}$ \cite{Rapp:1999ej,Coquet:2021lca}. They can therefore provide valuable insight into the entire time evolution and dynamics of the hot system. However, measurements of the thermal photon signal in heavy-ion collisions face some considerable challenges, such as a small production cross section due to the electromagnetic coupling and significant combinatorial and physical backgrounds from hadron decays.

In proton-proton (pp) collisions, the di-electron low and intermediate mass spectrum can be well described by a combination of expected hadronic sources, the so-called cocktail\,\cite{PHENIX:2008qav, PHENIX:2009gyd, STAR:2012dzw, ALICE:2018fvj, ALICE:2018gev}. The low-mass region (LMR, $m_{ee} < \SI{1.1}{\giga\electronvolt} $) is dominated by light-flavoured vector meson decays, whereas the intermediate-mass region (IMR, $\SI{1.1}{\giga\electronvolt} < m_{ee} < \SI{2.7}{\giga\electronvolt}$) receives contributions mostly from semileptonic decays of charm and bottom hadrons correlated through flavour conservation. Despite the overwhelming backgrounds, significant enhancements of direct real or low-mass virtual photons have been observed in heavy-ion collisions at the SPS, RHIC and LHC\,\cite{CERES:1995vll, CERESNA45:1997tgc, CERES:2006wcq, CERESNA45:2002gnc, NA60:2006ymb, NA60:2008ctj, PHENIX:2008uif, PHENIX:2009gyd, STAR:2013pwb, STAR:2015tnn, PHENIX:2014nkk, ALICE:2015xmh}. The enhancements are well described by exponential distributions in $p_{T,\gamma}$ and are thus indicative of thermal radiation consistent with average (effective) temperatures in the range of $200-\SI{300}{\mega\electronvolt}$\,\cite{PHENIX:2008uif, PHENIX:2014nkk, ALICE:2015xmh}. In the intermediate-mass window, an excess over decay di-electrons has so far been observed only at the SPS\,\cite{NA60:2006ymb, NA60:2008ctj, NA60:2008dcb, NA60:2007lzy}. In order to discern the characteristics of the subtle thermal photon signal in this range at the LHC, that promises increased sensitivity to QGP thermal radiation, it is crucial to understand the di-electron yield originating from the hard production of photons as well as from heavy-flavour particle decays.

In this publication, we report on theoretical progress in the description of the di-electron mass spectrum in the intermediate-mass region in setups similar to those used in ALICE experimental analyses. First, we calculate the contribution to the {\em direct} photon production signal originating from the hard scattering. This component is usually extracted from the data\,\cite{ALICE:2018ael}, assuming the invariant mass shape from the Kroll-Wada prescription\,\cite{Kroll:1955zu} with additional constraints from real direct photon measurements, and must be subtracted from the total direct photon signal for reliable temperature estimates\,\cite{Klasen:2013mga,ALICE:2015xmh}, especially in the intermediate transverse momentum region. Here, for the first time, we obtain a prediction for the di-electron invariant mass spectrum based on a first-principles calculation using the Drell-Yan process in association with an extra jet at next-to-leading order (NLO) QCD accuracy matched to parton showers. Second, we perform a thorough analysis of the dominant {\em decay} backgrounds with a variation of perturbative scales and parton distribution functions (PDFs), also taking into account cold nuclear effects through nuclear PDFs. Such uncertainties can have an appreciable impact on the systematic uncertainty estimates in the measured di-electron spectrum baseline in pp as well as heavy-ion collisions. Studying this process is also of broader interest, as it offers a means to test perturbative QCD (pQCD) calculations and Monte Carlo (MC) event generators at the limits of perturbativity. In heavy-ion collisions it can be used in studies of energy loss, levels of thermalisation of charm and bottom quarks within the medium, and mechanisms of heavy-quark hadronisation\,\cite{Greco:2003vf,Andronic:2003zv,Baier:1996sk,Braaten:1991we}.
The distribution of correlated $e^+e^-$ pairs stemming from charm-hadron decays provides information about kinematic correlations between charm and anticharm quarks, shedding light on the production mechanisms and offering sensitivity to soft heavy-flavour production\,\cite{LHCb:2015swx,CMS:2021lab,ALICE:2021mgk} and thermalization, in PbPb collisions.

The manuscript is organised as follows. We first present general theoretical considerations and define our signal and background in the di-electron observables in the kinematic region of interest in \Cref{sec:signalAndBackground}. 
We then explain our simulation setup including our workflow and the event selection in \Cref{sec:sumulationSetup}. 
In \Cref{sec:predictions} we present our results for virtual photon and heavy flavour production combined into a new prediction for the ``hadronic cocktail''. 
Finally, we summarise this work in \Cref{sec:concl}.

\section{Signal and background}
\label{sec:signalAndBackground}

The intermediate mass window region in measurements of di-electron spectra is characterised by a lepton pair with a relatively low invariant mass, between roughly 1 and 3 GeV, as well as relatively low transverse momentum, below 10 GeV. Besides the thermal photon signal, there are two dominant production modes in this kinematic regime.  One of them is the direct photon production accompanied by a photon conversion into di-electrons. In the following we will refer to this contribution as the {\em signal}. The second contribution is the heavy-flavoured hadron production followed by a decay into a di-electron pair.\footnote{The light-flavoured hadron production with di-electron decay dominates the invariant mass spectrum below 1 GeV, but is rather suppressed above.} This contribution, instead, we will refer to as the {\em background}. 

The signal is the direct production of photons with non-zero virtuality that decay into di-electrons, that is the Drell-Yan (DY) process. The tree-level DY process is described by the partonic process $q\bar q \to \gamma^* \to l^+l^-$ and is of the order $\mathcal{O}(\alpha^2)$ in the perturbative expansion of the electromagnetic coupling constant. In the absence of any other particles the $e^+ e^-$ pair is produced back-to-back and the system carries no transverse momentum, falling out of the kinematic region of our interest. To get around this problem, one must produce a pair of di-electrons in association with one jet at order $\mathcal{O}(\alpha^2\alpha_S)$ in the simultaneous expansions of the electromagnetic and strong couplings constants. At this order one does not only receive contributions from the $q\bar q$ channel, but also from the $qg$ channel and therefore picks up sensitivity to the gluon density.\footnote{Furthermore, the NLO QCD corrections, $\mathcal{O}(\alpha^2\alpha_S^2)$, can also be gluon-gluon initiated.} Note that this process also receives contributions from the topology with the photon replaced by the $Z$ boson and the corresponding $\gamma-Z$ interference, which we both include, but their contributions are negligible in the invariant-mass regime far below the $Z$-boson mass. We simulate our signal at NLO QCD accuracy, i.e.~including higher order real and virtual corrections up to $\mathcal{O}(\alpha^2\alpha_S^2)$, using the \texttt{Zj} process \cite{Alioli:2010qp} from the \texttt{POWHEG\,BOX\,V2} package \cite{Nason:2004rx, Frixione:2007vw, Alioli:2010xd}, which is now also available in Multi-Scale improved NLO (MiNLO) \cite{Hamilton:2012np} and NNLO (MiNNLO) \cite{Monni:2019whf,Monni:2020nks} accuracies.

In the limit of zero di-electron virtuality one recovers real direct photon production at $\mathcal{O}(\alpha\alpha_S)$, in which the photon is produced on-shell and does not decay into di-electrons. Similarly to the fragmentation of jets into photons in direct real photon production, contributions from the fragmentation of jets into di-electrons are possible. In analogy to our calculation of real direct photons \cite{Jezo:2016ypn} within the POWHEG framework, the {\tt Zj} process includes this fragmentation contribution, partly generated through parton shower emissions of the photon.\footnote{Contrary to real photons, in virtual photon production there is no explicit divergence that must be absorbed into fragmentation functions and, instead, the divergence is regulated by the photon virtuality \cite{Klasen:1997jm,Klasen:2002xb}. Nevertheless, the resulting logarithms $\ln(m_{ee}/p_{T,ee})$ arising per order of $\alpha_S$ could in principle become large and spoil the convergence of the perturbative series. We keep track of the $m_{ee}/p_{T,ee}$ ratio and find it ranges from $10^{-1}$ to $10^{1}$ with a sharp peak around 1. We thus expect our predictions to be stable with respect to such higher order effects in the kinematic region of our interest.}

We now move on to the discussion of the background, which can be further split into the open production of charm and bottom quarks \cite{Rapp:1999ej,Mohanty:2013rna}. While photon-induced or DY-like production of heavy quarks is possible, due to the ratio of couplings strengths $\alpha^2/\alpha_S^2\lesssim\SI{0.5}{\%}$ it is typically sufficient to only consider the QCD production \cite{Bonciani:2015hgv, Pagani:2016caq}. That is, heavy-flavour partons are only produced from a subset of the QCD Feynman diagrams of light-flavour production, namely where the heavy quarks in the final state are connected by a fermion line and thus belong to the same flavour generation independently of the initial-state partons. The heavy quarks then hadronise predominantly into $D$ and $B$ mesons, which further decay, in a correlated manner, into electrons. This results in correlated electron pairs stemming from two different hadrons originating from the same hard process. To simulate our background at NLO QCD accuracy ($\mathcal{O}(\alpha_S^{3})$), we use the \texttt{hvq} process \cite{Frixione:2007nw} from \texttt{POWHEG\,BOX\,V2}. 

Both simulations, of the signal and of the background, are further interfaced to \texttt{Pythia8}. In the case of the signal, it only provides higher order QCD and QED corrections in the soft-collinear approximation via parton showers, whereas in the case of the background we rely on it also for the hadronisation and the heavy-flavoured hadron decays. We use \texttt{Pythia8} version 8.3.08 \cite{Bierlich:2022pfr}.

\section{Simulation setup and event selection}
\label{sec:sumulationSetup}

In this manuscript we focus on the production of di-electrons, open charm and bottom pairs in proton-proton (pp), proton-lead (pPb) and lead-lead (PbPb) collisions at the LHC with $\sqrt{s} = 5.02$ TeV (per nucleon pair). The experimental pp and pPb data from the corresponding ALICE measurement \cite{ALICE:2020mfy} are  available in {\tt HEPdata} \cite{Maguire:2017ypu}. The
ALICE data in PbPb have been recently published \cite{ALICE:2023jef}, but the numerical values are not yet available in {\tt HEPdata}.
Unless stated otherwise, we use MSHT20nlo\,\cite{Bailey:2020ooq} PDFs for the proton and nCTEQ15HQ\,\cite{Duwentaster:2022kpv} nuclear PDFs for lead as provided by \texttt{LHAPDF6}\,\cite{LHAPDF} with the corresponding value of $\alpha_S(M_Z)$.

We generate the Les Houches Event (LHE) samples using {\tt POWHEG\,BOX\,V2} and shower them using {\tt Pythia8} piloted by the {\tt main-PYTHIA83-lhef} executable, which is in turn coupled to {\tt Rivet\,3.1.7} \cite{RIVET} based on the {\tt Rivet\,2} interface originally developed in Ref.\,\cite{Jezo:2018yaf} and later upgraded for {\tt Rivet\,3} in Ref.\,\cite{FerrarioRavasio:2023kjq}. The relevant physics input parameters are chosen from Ref.~\cite{ParticleDataGroup:2022pth} with
\begin{align*} 
   \alpha & =  1/128.89, & m_c & = \SI{1.27}{\giga\electronvolt}, & m_b & = \SI{4.18}{\giga\electronvolt}.
\end{align*}
For the production of heavy quarks, the  renormalisation and factorisation scales are chosen dynamically as $\mu_r = \mu_f =\sqrt{p_T^2 + m_q^2}$, where $p_T \neq p_{T,ee}$ is the transverse momentum of the heavy quark and $q$ labels its flavour. For $c\bar c$ a three-flavour scheme is used, while for $b\bar b$ we use a four-flavour scheme. The corresponding running of $\alpha_S$ is calculated with internal {\tt POWHEG} subroutines.
For the Drell-Yan process instead we set the value of the renormalisation and factorisation scales to $\mu_r = \mu_f = \max\left(2\ {\rm GeV},\sqrt{p_{T,\gamma}^2+p_{\gamma}^2}\right)$ with its minimum value frozen at 2 GeV and use a five-flavour scheme.
In both cases we calculate perturbative scale uncertainties using the standard factor-two seven-point $\mu_r, \mu_f$ scale variation method, excluding relative factors of four. 

In {\tt Pythia8}, we leave parton shower, hadronisation and MPI enabled.
Hadron decays are only enabled for our background predictions, as to avoid decay photons in our signal. 
Analogously, in heavy flavour production we disable the QED shower, but leave it on for the production of virtual photons in order to generate the fragmentation contribution. 
To match to POWHEG we employ the standard procedure in which shower evolution starts unrestricted but shower emissions are vetoed using {\tt PowhegHooks}.
In \Cref{tab:pythia} we list values of selected \texttt{Pythia8} parameters for each process.
\begin{table}
\begin{center}
\caption{
    Parameters used in \texttt{Pythia} for the different processes. 
    Values differing from the default are highlighted in bold.
}
\label{tab:pythia}
\begin{tabular}{ c | c | c }
 {\tt Pythia} setting                   & {\tt Zj}       & {\tt hvq} \\ \hline\hline 
 \verb|PartonLevel:ISR             |&      on    &      on  \\
 \verb|PartonLevel:FSR             |&      on    &      on  \\
 \verb|PartonLevel:MPI             |&      on    &      on  \\
 \verb|PartonLevel:Remnants        |&      on    &      on  \\ \hline
 \verb|PartonShowers:model         |&      1     &      1   \\ \hline
 \verb|SpaceShower:pTmaxMatch      |&\bf     2   &\bf     2 \\
 \verb|TimeShower:pTmaxMatch       |&\bf     2   &\bf     2 \\
 \verb|POWHEG:veto                 |&\bf     1   &\bf     1 \\
 \verb|POWHEG:pTdef                |&\bf     1   &\bf     1 \\ \hline
 \verb|SpaceShower:QEDshowerByL    |&   on   &\bf   off \\
 \verb|SpaceShower:QEDshowerByQ    |&   on   &\bf   off \\
 \verb|TimeShower:QEDshowerByGamma |&   on   &\bf   off \\
 \verb|TimeShower:QEDshowerByL     |&   on   &\bf   off \\
 \verb|TimeShower:QEDshowerByOther |&   on   &\bf   off \\
 \verb|TimeShower:QEDshowerByQ     |&   on   &\bf   off \\ \hline
 \verb|HadronLevel:Hadronize       |&      on    &       on \\
 \verb|HadronLevel:Decay           |&\bf   off   &       on \\
\end{tabular}
\end{center}
\end{table}

\begin{figure}
    \centering
    \begin{subfigure}[b]{0.48\textwidth}
    \centering
    \includegraphics[width=0.85\textwidth]{{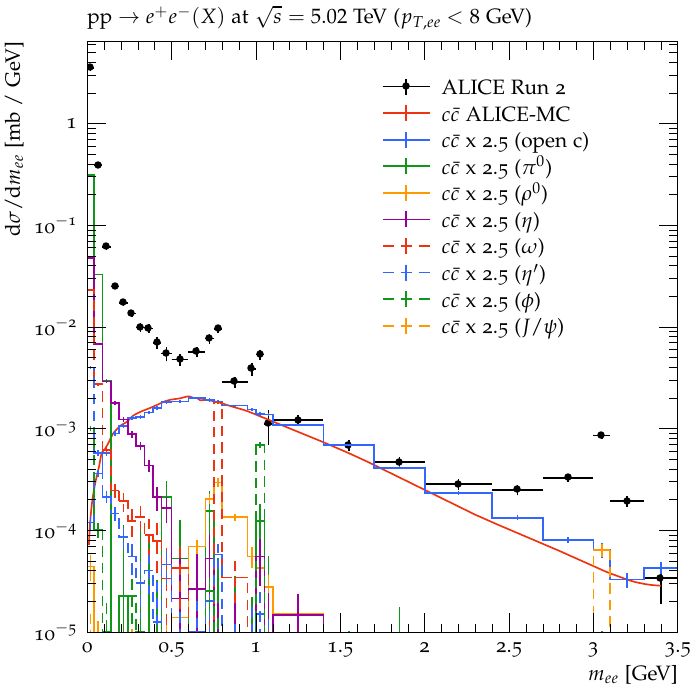}}
    \end{subfigure}
    \begin{subfigure}[b]{0.48\textwidth}
    \centering
    \includegraphics[width=0.85\textwidth]{{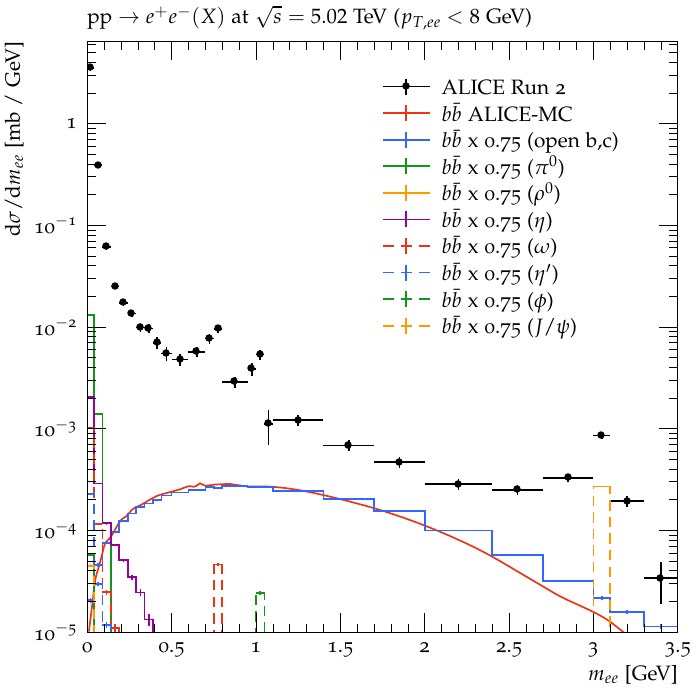}}
    \end{subfigure}
    \caption{
        Di-electron mass spectra for $\text{pp}\to c\bar c(X)$ (left) and $\text{pp}\to b\bar b(X)$ (right) production, decomposed into open heavy flavour continuum and the resonance production in the LMR and IMR. All the contributions were normalised as in~\cite{ALICE:2020mfy}.
        \
    }
    \label{fig:decomp_pp}
\end{figure}

Our event selection and histogramming of the observables is performed by a {\tt Rivet} analysis prepared for this publication that mimics the analysis of Ref.~\cite{ALICE:2020mfy} and which we include in the ancillary files.
We select all electrons and positrons that fulfil the acceptance cuts of $|\eta_{e}| < 0.8$, required due to ALICE's central barrel geometry, and $\SI{0.2}{GeV} < p_{T,e} < \SI{10}{GeV}$.
We then use the electron and positron to build the transverse momentum and virtuality of the pair, $p_{T,ee}$ and $m_{ee}$.
As in the ALICE analysis, we consider the transverse momentum and invariant mass ranges $\SI{0}{GeV}< p_{T,ee} < 8$ GeV and $\SI{0.5}{GeV}<m_{ee} <\SI{1.1}{GeV}$ or $\SI{1.1}{GeV}<m_{ee} <\SI{2.7}{GeV}$ for the di-electron pair. For the invariant mass spectra, we also consider the range $\SI{0}{GeV}< m_{ee} <\SI{7}{GeV}$, which extends upon the ALICE measurement that is only available in the range $\SI{0}{GeV}< m_{ee} <\SI{3.5}{GeV}$. 

We remove the combinatorial background using the same-sign approximation, as in Ref.~\cite{ALICE:2018fvj}, where the raw pair signal $S$ is obtained with the formula
\begin{equation}
    S = \underbrace{\left( N_{+-} \right)}_{\text{OS}} - R_{\text{acc}} \underbrace{\left( N_{++} + N_{--} \right)}_{\text{SS}}\ ,
\end{equation}
where we estimate the number of same-sign (SS) pairs with an arithmetic mean, which is more suitable than a geometric mean when considering a contribution to the spectrum due to a single process, and where we set the relative acceptance correction factor, $R_\text{acc}$, equal to one. Moreover $N_{+-}$ is the number of opposite-sign (OS) pairs and $N_{++}$, $N_{--}$ the numbers of positron-positron, electron-electron pairs in the signal region. 
Assuming that the SS pairs are uncorrelated, the subtracted term will correspond to the uncorrelated OS pairs at the high statistical limit.
This gives the raw pair signal $S$ with unwanted uncorrelated contributions removed.

In order to isolate the continuum contribution to open heavy flavour production from the heavy quark {\tt POWHEG+Pythia8} sample we classify electrons according to their origin.
In \Cref{fig:decomp_pp}, we show di-electron invariant mass spectra from $\text{pp}\to c\bar c(X)$ and $\text{pp}\to b\bar b(X)$ production. 
In both cases, below 1 GeV, the spectrum receives contributions from heavy to light flavoured hadron cascade decays: $\pi^0, \rho^0, \eta, \omega, \eta^\prime, \phi$. 
Above 1 GeV, only the $J/\psi$ peak is visible at around 3 GeV, whereas in between, in the IMR, the continuum dominates. 

This inspires our definition of the continuum contribution to the hadronic cocktail as the {\tt POWHEG+Pythia8} subsample with electrons from light hadron and $J/\psi$ decays filtered away. 
Practically speaking, we require all leptons to have a parent particle that is either a $b$ or $c$ flavoured hadron by inspecting the event record in {\tt Rivet}.
It is worth noting that this implies that the hadron decay chains $b \to c \to e$ are not filtered away.
This is meant to align with the definition used in ALICE in which contributions removed by filtering based on origin are instead obtained using the Monte Carlo event generator EXODUS\,\cite{PHENIX:2009gyd}.
Our predictions, blue histograms, reproduce the predictions from ALICE extracted from Ref.~\cite{ALICE:2020mfy}, $c\bar{c}(b\bar{b})$ ALICE-MC, which were obtained with \texttt{POWHEG+Pythia6} and CTEQ6.6 PDFs, red curves, very well.
We note that our predictions were normalised such that the area under the curve of our ``open c'' and ``open b'' prediction match the area under the ``$c\bar{c}$ ALICE-MC'' and ``$b\bar{b}$ ALICE-MC'' curves, respectively.

\section{Our predictions}
\label{sec:predictions}

In Refs.~\cite{ALICE:2020mfy,ALICE:2018fvj,ALICE:2018gev} the production of di-electron pairs ($e^+e^-$) in proton-proton collisions at $\sqrt{s} = (5.02, 7, 13)$ TeV was measured using the ALICE detector at the LHC.
The studies focused on the invariant mass and transverse momentum distributions of the $e^+e^-$ pairs.
The data was compared to a hadronic cocktail composed of expected di-electron distributions of known three dominant hadronic sources: light flavour decays, the $J/\psi$ resonance, and open heavy flavour decays.
Good agreement between data and the hadronic cocktail was observed over the entire mass range ($m_{ee} < 3.5$ GeV).
However, the extracted total cross sections for charm and beauty strongly depend on details of theoretical modelling (e.g.~\texttt{Pythia6} vs.~\texttt{POWHEG+Pythia6}). 
The same measurement was also carried out in proton-lead collisions at $\sqrt{s_{\text{NN}}} = 5.02$ TeV \cite{ALICE:2020mfy}.
Again good agreement with the cocktail was found assuming heavy flavour cross sections scale with the atomic mass number $A$ of the lead nucleus in pPb-collisions with respect to the pp reference.
The measured ratio $R_\text{pPb}$, corrected for such $A$ scaling, is consistent with unity within uncertainties in the IMR where most of the $e^+e^-$ pairs originate from correlated open heavy flavour hadron decays.
Thus, the uncertainties on the measured $p_{T,ee}$ dependence of $R_\text{pPb}$ are still too large to draw conclusions on the nuclear modification of heavy flavour production in pPb collisions.
In PbPb collisions, ALICE measured the $e^+e^-$ spectra at $\sqrt{s_{\text{NN}}} = (2.76, 5.02)$ TeV \cite{ALICE:2018ael, ALICE:2023jef}, but the former measurement is statistically limited.
Also here the cocktail with an additional scaling by $R^{b,c\to e}_{\text{PbPb}}$ due to final state medium effects is in reasonable agreement with the data, despite neglecting the cold nuclear matter effects.

In this section we study two predictions that enter those analyses. 
First we present our predictions for the prompt component of the di-electron spectrum due to direct virtual photon production in \Cref{sec:virtualPhotons}. 
This contribution has so far not been considered in the ALICE measurements.
Second we show our predictions for the open charm and bottom production, including a careful analysis of uncertainties from scale and PDF variations in \Cref{sec:openHeavyFlavour}. 
In \Cref{sec:combination} we analyse all our predictions together and assemble them into a new prediction for the hadronic cocktail with realistic uncertainties. 
Finally, \Cref{sec:leadLead} is dedicated to our predictions for PbPb collisions and a discussion on initial- and final-state matter effects. 

\subsection{The signal: direct virtual photon production \label{sec:virtualPhotons}}

\begin{figure}
    \centering
    \begin{subfigure}[b]{0.48\textwidth}
        \centering
        \includegraphics[width=0.85\textwidth]{{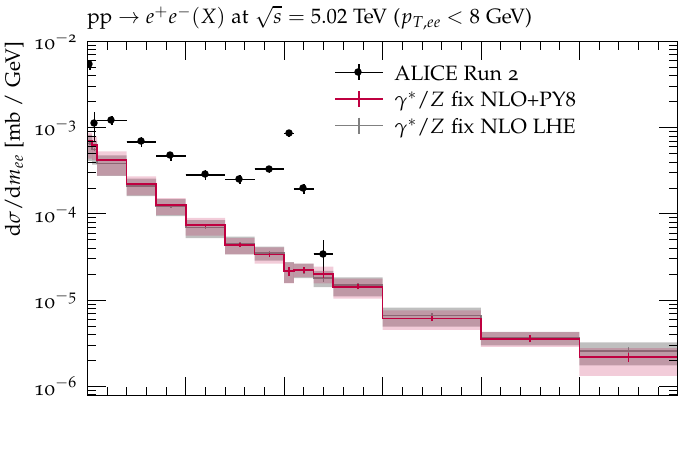}}
        \vspace{-2.30em}

        \includegraphics[width=0.85\textwidth]{{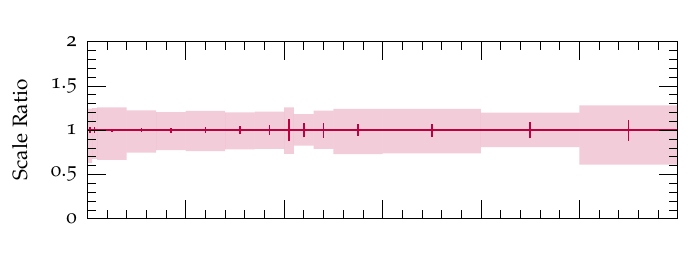}}
        \vspace{-2.30em}

        \includegraphics[width=0.85\textwidth]{{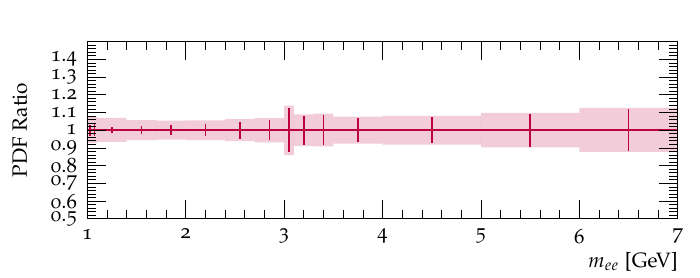}}
        \vspace{-2.30em}
    \end{subfigure}
    \hfill
    \begin{subfigure}[b]{0.48\textwidth}
        \centering
        \includegraphics[width=0.85\textwidth]{{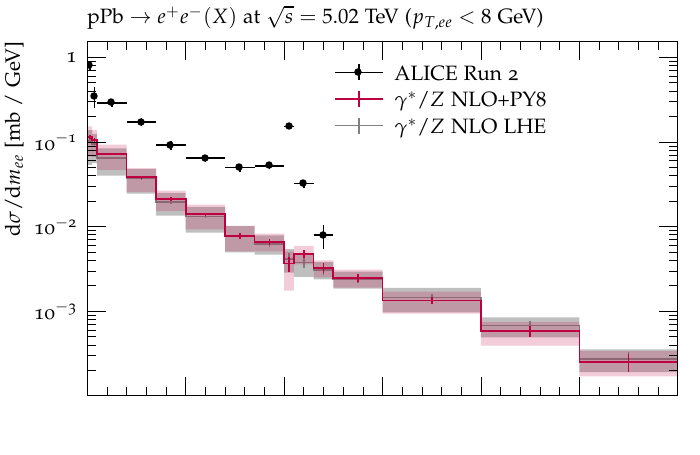}}
        \vspace{-2.30em}

        \includegraphics[width=0.85\textwidth]{{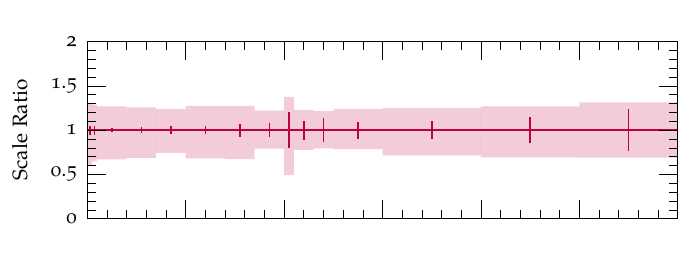}}
        \vspace{-2.30em}
	    
        \includegraphics[width=0.85\textwidth]{{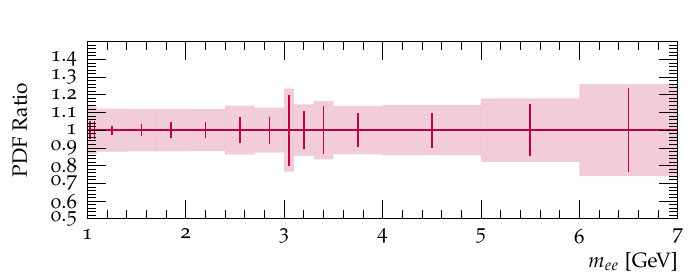}}
	\vspace{-2.30em}
    \end{subfigure}
    \vspace{3em}
    \caption{
        Di-electron pair invariant mass spectrum from virtual photon production in association with at least one jet at NLO+PS for pp (left) and pPb (right) obtained with \texttt{POWHEG+Pythia8} using MSHT20nlo and nCTEQ15HQ PDFs for p and Pb.
        The uncertainty band combines scale and PDF uncertainties in quadrature.
        The second and third ratio panels respectively show the relative  scale and PDF uncertainties independently.
        \label{fig:dy_NLO_inv}
    }
\end{figure}

\begin{figure}
    \centering
    \begin{subfigure}[b]{0.48\textwidth}
        \centering
        \includegraphics[width=0.85\textwidth]{{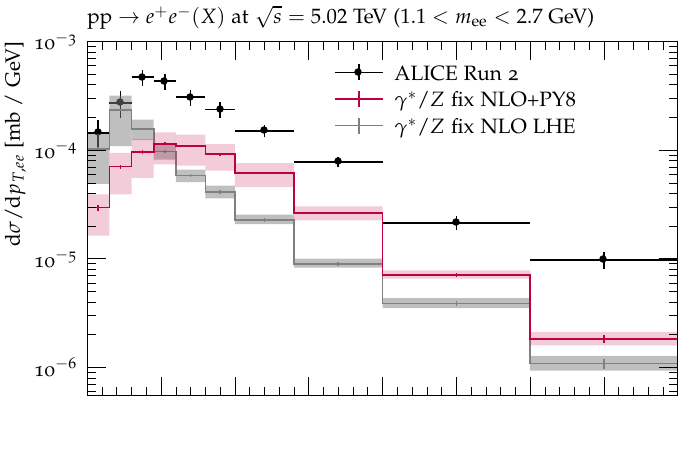}}
        \vspace{-2.30em}

        \includegraphics[width=0.85\textwidth]{{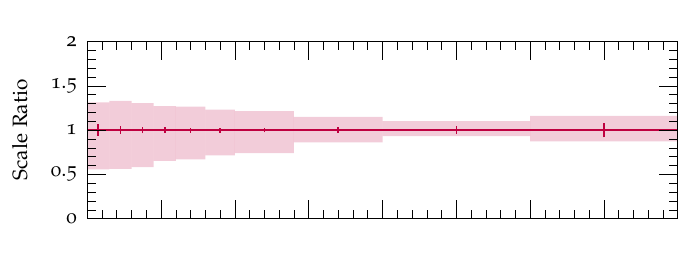}}
        \vspace{-2.30em}

        \includegraphics[width=0.85\textwidth]{{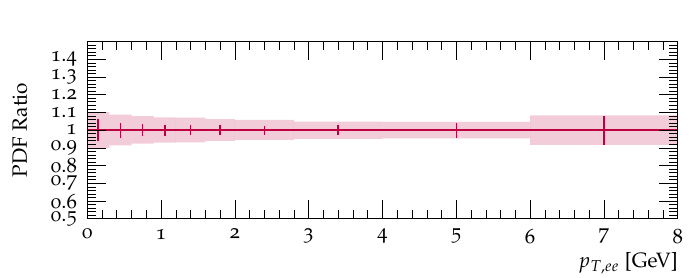}}
        \vspace{-2.30em}
    \end{subfigure}
    \hfill
    \begin{subfigure}[b]{0.48\textwidth}
        \centering
        \includegraphics[width=0.85\textwidth]{{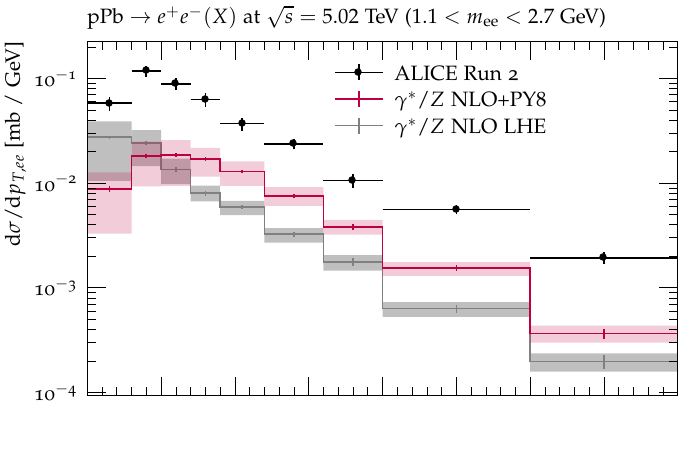}}
        \vspace{-2.30em}

        \includegraphics[width=0.85\textwidth]{{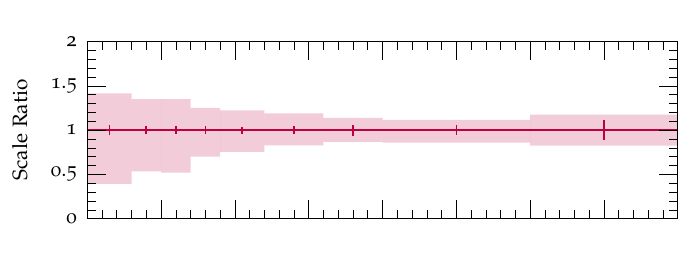}}
        \vspace{-2.30em}

        \includegraphics[width=0.85\textwidth]{{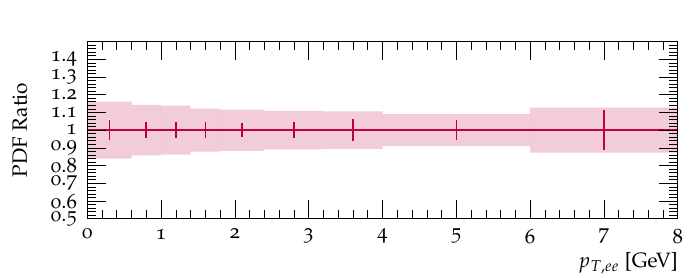}}
        \vspace{-2.30em}
    \end{subfigure}
    \vspace{3em}
    \caption{
        Di-electron pair transverse momentum spectrum from virtual photon production in association with at least one jet at NLO+PS for pp (left) an pPb (right) obtained with \texttt{POWHEG+Pythia8} using MSHT20nlo and nCTEQ15HQ PDFs for as p and Pb.
        The uncertainty band combines scale and PDF uncertainties in quadrature.
        The second and third ratio panels respectively show the relative scale and PDF uncertainties independently.
        \label{fig:dy_NLO_pt}
    }
\end{figure}

\Cref{fig:dy_NLO_inv} shows the invariant mass spectra of the di-electron system originating from virtual photons produced in association with at least one jet in 5.02 TeV pp and pPb in collisions at
NLO+PS in red and at the NLO LHE level\footnote{NLO LHE corresponds to an NLO prediction supplemented by the Sudakov form factor of the NLO emission.} in grey.
Going below an invariant mass of $m_{ee}=1$ GeV, the cross section becomes less reliable due to the photon pole at \mbox{$m_{ee} = 0$}.
It is therefore not surprising that the invariant mass spectrum increases continuously towards lower masses, see the top panels.
The total uncertainties combined in quadrature are dominated by the variation of scales and are about $\pm$\SI{25}{\%} across the whole invariant mass range, but up to $\pm$\SI{50}{\%} in the low tail of the transverse momentum spectra (see below). 
In contrast, the PDF uncertainties are roughly $\pm$\SI{10}{\%} for proton-proton collisions and only a few percent more for proton-lead, see the third and the fourth panels.
Note that the PDF uncertainty band beyond $m_{ee} > 3$ GeV,  saturated by the Monte Carlo errors due to limited size of our event sample, likely overestimates the true uncertainty.
The difference between the sizes of the cross section of the different initial states is a relative increase by $A=208$ from pp to pPb.
For the $p_T$ spectrum in \Cref{fig:dy_NLO_pt} we observe that the parton shower emissions beyond $\mathcal{O}(\alpha_S)$ soften the low tail and harden the high tail of the $p_{T,ee}$ spectrum (NLO LHE vs.\,NLO+Py8). 
Such a transformation into a more physical distribution is expected from matching to parton showers.

The virtual photon production is a subleading but not a negligible contribution to the di-electron pair production in this kinematic regime. 
It reaches fractions up to 45\% in the range of 1.2 to 3 GeV of the invariant mass spectrum, where the depletion at 3 GeV is due to the $J/\psi$ contribution in the measured spectrum. 
Similarly, in the transverse momentum spectrum it contributes up to 40\% in the intermediate $p_T$ range of 1 to 4 GeV. 
Note that our scale choice, appropriate for this kinematic regime, enhances the virtual photon production cross section appreciably. 
With the (inappropriate) choice $\mu_f=\mu_r=m_Z$ this contribution would be only at about 5\% with very similarly shaped distributions.

\begin{figure}
    \centering
    \begin{subfigure}[b]{0.48\textwidth}
        \centering
        \includegraphics[width=0.85\textwidth]{{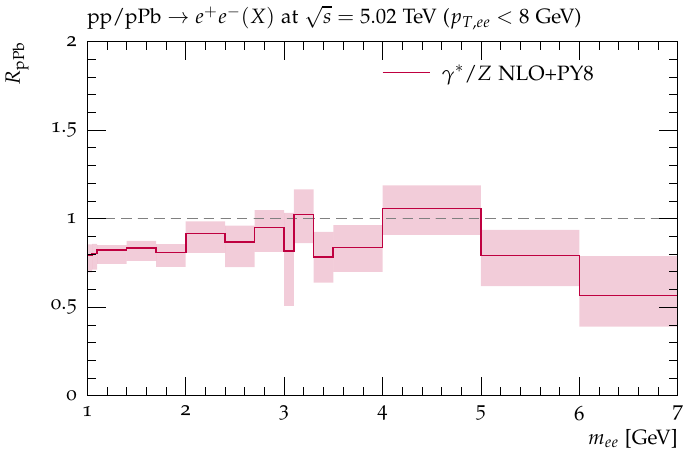}}
        \vspace{-2.30em}
    \end{subfigure}
    \hfill
    \begin{subfigure}[b]{0.48\textwidth}
        \centering
        \includegraphics[width=0.85\textwidth]{{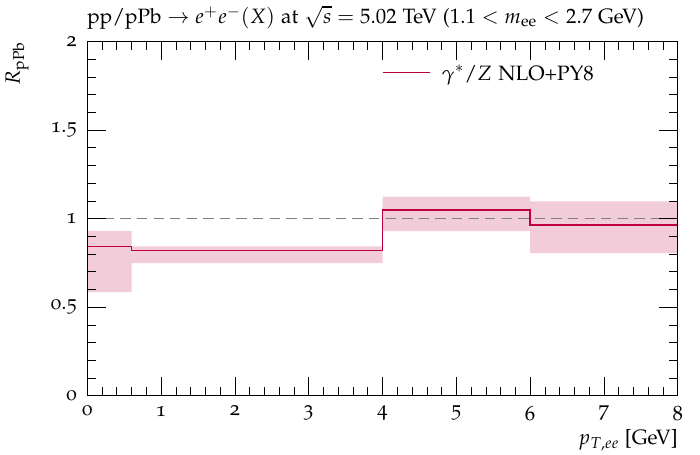}}
        \vspace{-2.30em}
    \end{subfigure}
    \vspace{3em}
    \caption{
        Di-electron pair ratios $R_\text{pPb}$ of the invariant mass (left) and transverse momentum distributions (right) for the Drell-Yan process.
        The uncertainty bands are correlated scale uncertainties. 
        \label{fig:dy_NLO_rpPb}
    }
\end{figure}

In \Cref{fig:dy_NLO_rpPb} we show the ratio of pPb and pp predictions for the virtual photon production. Both the invariant mass and the transverse momentum spectra lie below one by up to 20-30\% with an appreciable kinematic dependence. The uncertainty band represents the usual 7-point scale variation band obtained by taking the envelope of ratios of pPb over pp predictions with the same scale choice. Contrary to the usual expectation for the scale uncertainty to cancel out in the $R_{\text{pPb}}$ ratio, it remains relatively large at about 25\% across the whole $m_{ee}$ and $p_{T,ee}$ range as compared to roughly $\pm 15\%$ to $\pm 40\%$ in the absolute predictions.

\FloatBarrier

\subsection{The background: open charm and bottom production\label{sec:openHeavyFlavour}}

\begin{figure}
    \centering
    \begin{subfigure}[b]{0.48\textwidth}
        \centering
        \includegraphics[width=0.85\textwidth]{{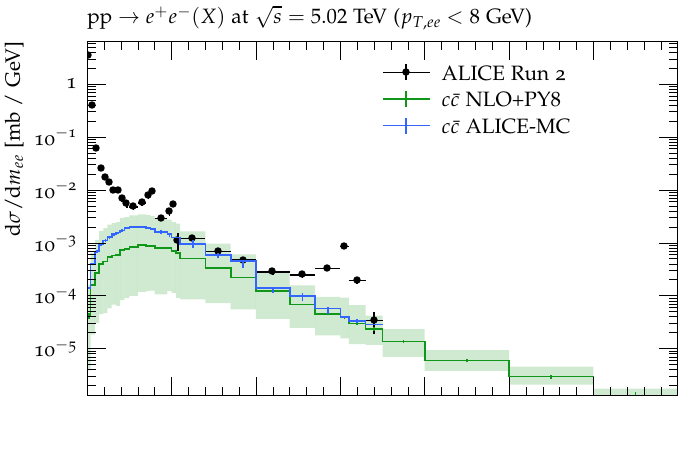}}
        \vspace{-2.30em}

        \includegraphics[width=0.85\textwidth]{{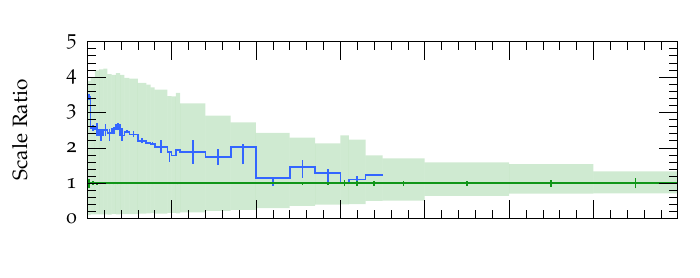}}
        \vspace{-2.30em}

        \includegraphics[width=0.85\textwidth]{{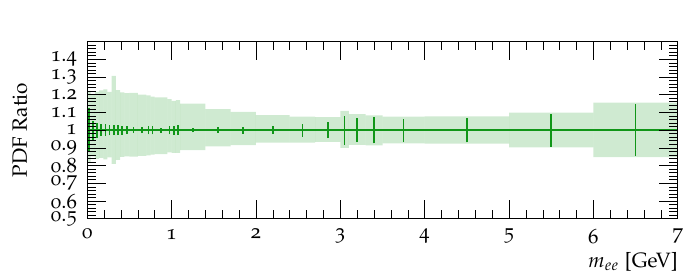}}
        \vspace{-2.30em}
    \end{subfigure}
    \hfill
    \begin{subfigure}[b]{0.48\textwidth}
        \centering
        \includegraphics[width=0.85\textwidth]{{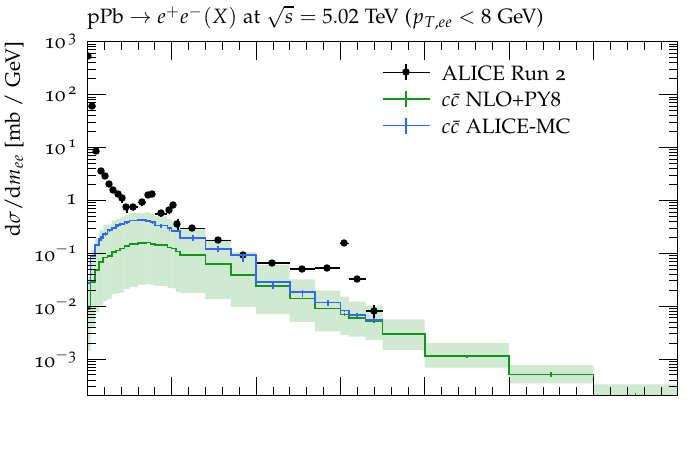}}
        \vspace{-2.30em}

        \includegraphics[width=0.85\textwidth]{{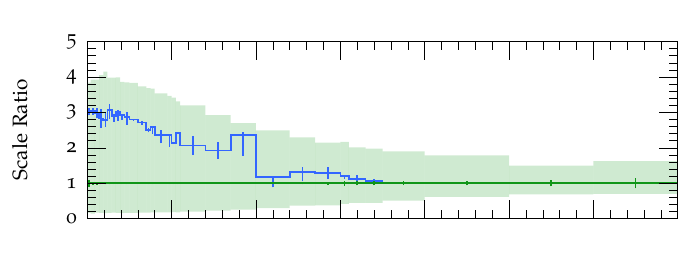}}
        \vspace{-2.30em}

        \includegraphics[width=0.85\textwidth]{{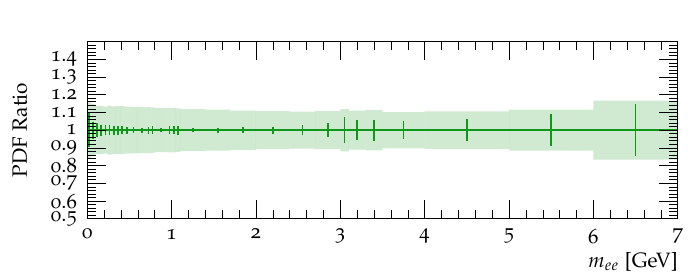}}
        \vspace{-2.30em}
    \end{subfigure}
    \vspace{3em}
    \caption{
        Invariant di-electron mass spectra in open charm production $\text{pp, pPb} \to c \bar c(X)$ obtained with \texttt{POWHEG+Py8} and MSHT20nlo and nCTEQ15HQ PDFs for p and Pb. 
        The second row shows the scale and the third row the PDF uncertainties. 
    }
    \label{fig:cc_inv}
\end{figure}

\begin{figure}
    \centering
    \begin{subfigure}[b]{0.48\textwidth}
        \centering
        \includegraphics[width=0.85\textwidth]{{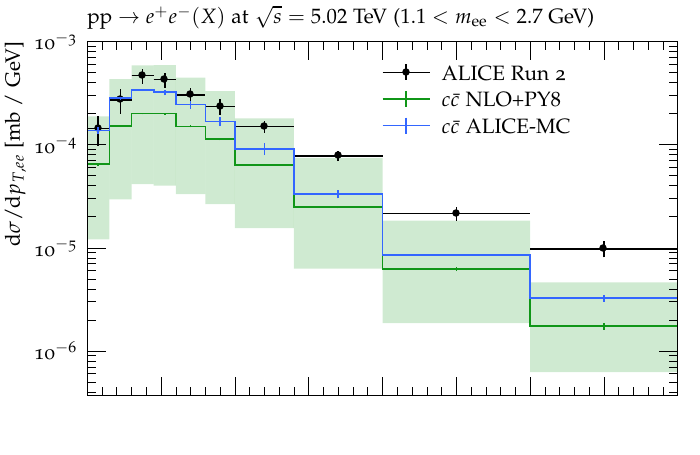}}
        \vspace{-2.30em}

        \includegraphics[width=0.85\textwidth]{{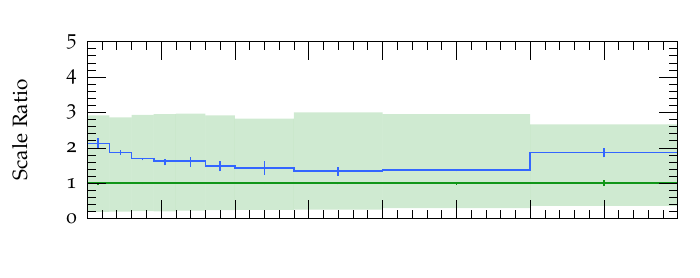}}
        \vspace{-2.30em}

        \includegraphics[width=0.85\textwidth]{{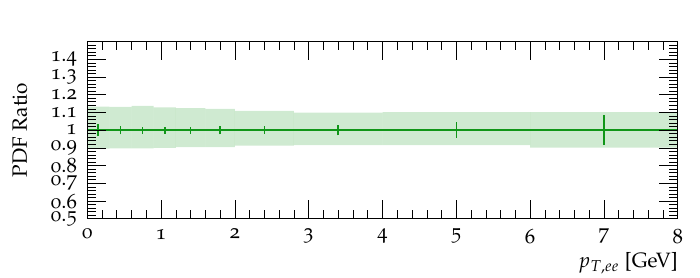}}
        \vspace{-2.30em}
    \end{subfigure}
    \hfill
    \begin{subfigure}[b]{0.48\textwidth}
        \centering
        \includegraphics[width=0.85\textwidth]{{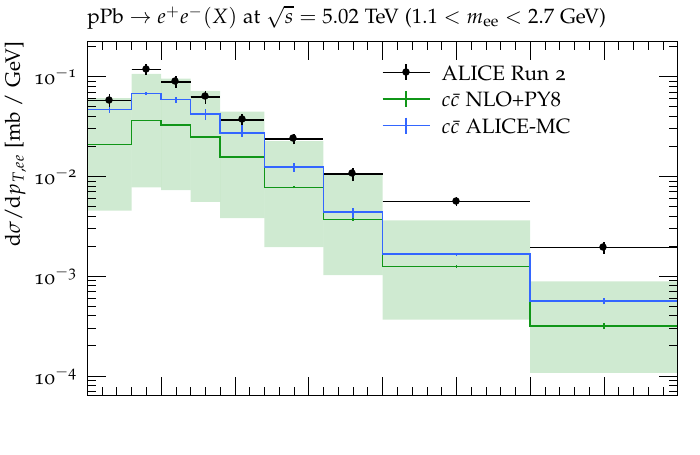}}
        \vspace{-2.30em}

        \includegraphics[width=0.85\textwidth]{{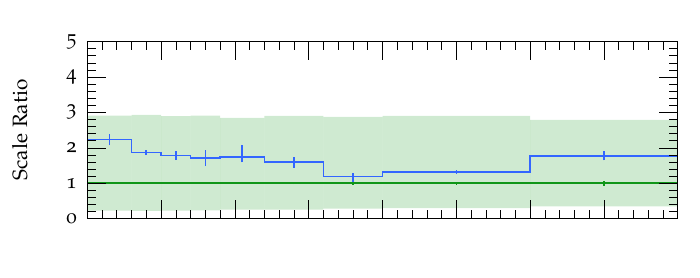}}
        \vspace{-2.30em}

        \includegraphics[width=0.85\textwidth]{{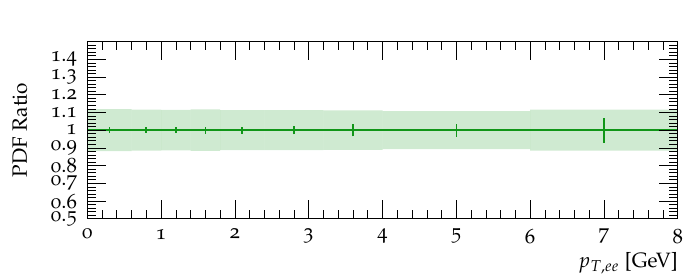}}
        \vspace{-2.30em}
    \end{subfigure}
    \vspace{3em}
    \caption{
        Di-electron transverse momentum spectra in open charm production $\text{pp, pPb} \to c \bar c(X)$ obtained with  \texttt{POWHEG+Py8} and MSHT20nlo and nCTEQ15HQ PDFs for p and Pb. 
        The second row shows the scale and the third row the PDF uncertainties. 
    }
    \label{fig:cc_imr}
\end{figure}

\begin{figure}
    \centering
    \begin{subfigure}[b]{0.48\textwidth}
        \centering
        \includegraphics[width=0.85\textwidth]{{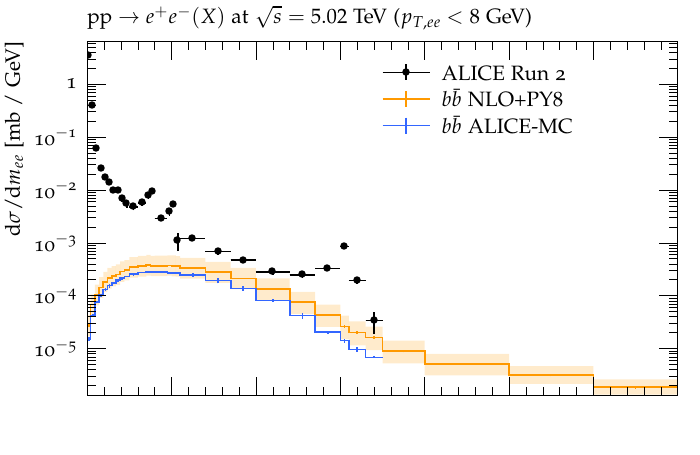}}
        \vspace{-2.30em}

        \includegraphics[width=0.85\textwidth]{{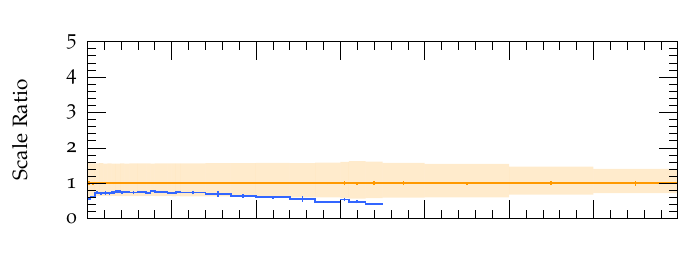}}
        \vspace{-2.30em}

        \includegraphics[width=0.85\textwidth]{{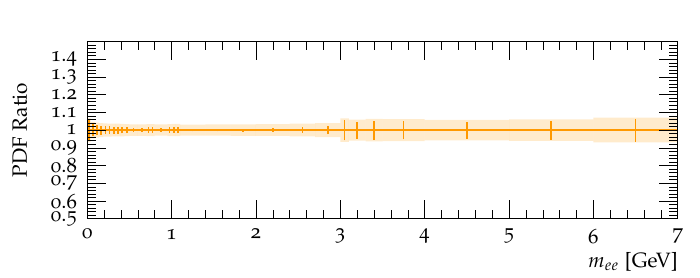}}
        \vspace{-2.30em}
    \end{subfigure}
    \hfill
    \begin{subfigure}[b]{0.48\textwidth}
        \centering
        \includegraphics[width=0.85\textwidth]{{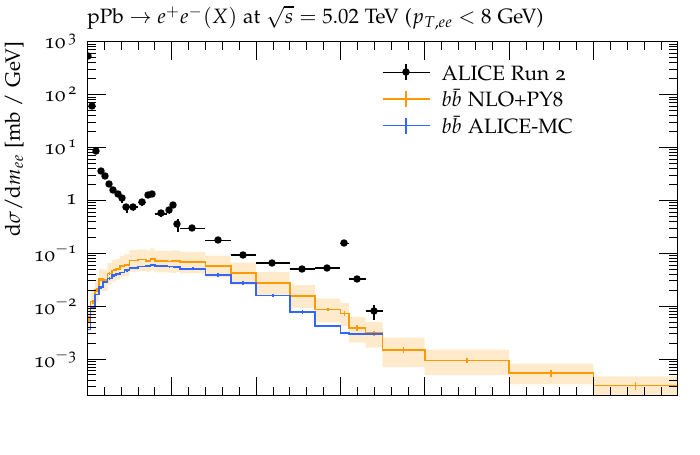}}
        \vspace{-2.30em}

        \includegraphics[width=0.85\textwidth]{{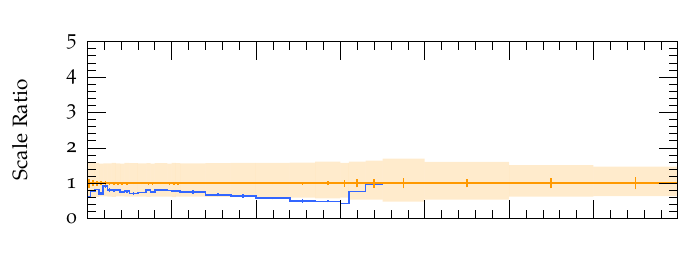}}
        \vspace{-2.30em}

        \includegraphics[width=0.85\textwidth]{{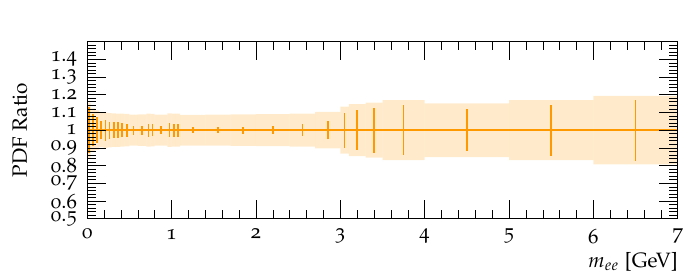}}
        \vspace{-2.30em}
    \end{subfigure}
    \vspace{3em}
    \caption{
        Invariant di-electron mass spectra in open bottom production $\text{pp, pPb} \to b \bar b(X)$ obtained with \texttt{POWHEG+Py8} and MSHT20nlo and nCTEQ15HQ PDFs for p and Pb. 
        The second row shows the scale and the third row the PDF uncertainties. 
   }
    \label{fig:bb_inv}
\end{figure}

\begin{figure}
    \centering
    \begin{subfigure}[b]{0.48\textwidth}
        \centering
        \includegraphics[width=0.85\textwidth]{{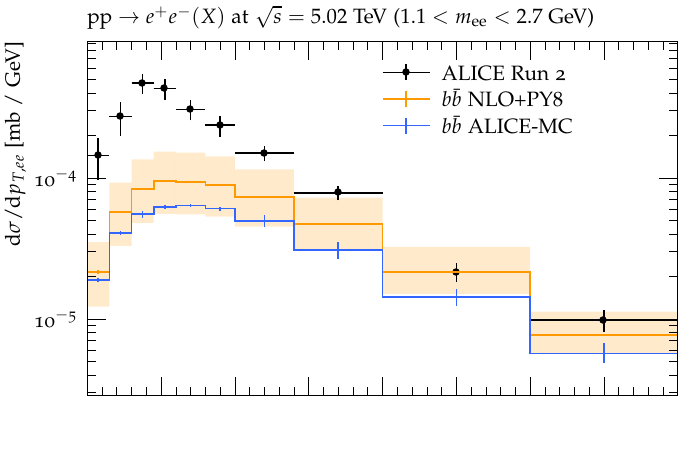}}
        \vspace{-2.30em}

        \includegraphics[width=0.85\textwidth]{{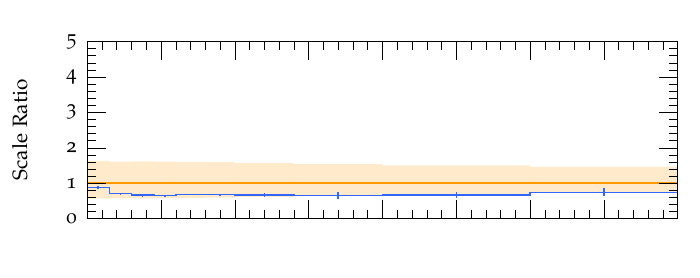}}
        \vspace{-2.30em}

        \includegraphics[width=0.85\textwidth]{{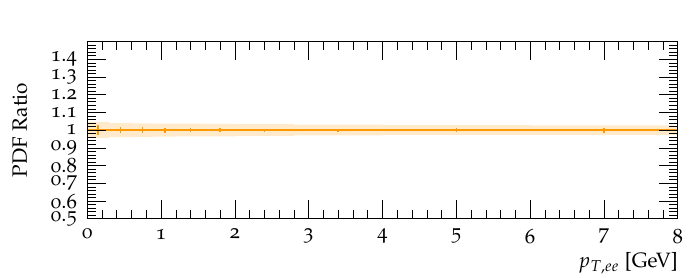}}
        \vspace{-2.30em}
    \end{subfigure}
    \hfill
    \begin{subfigure}[b]{0.48\textwidth}
        \centering
        \includegraphics[width=0.85\textwidth]{{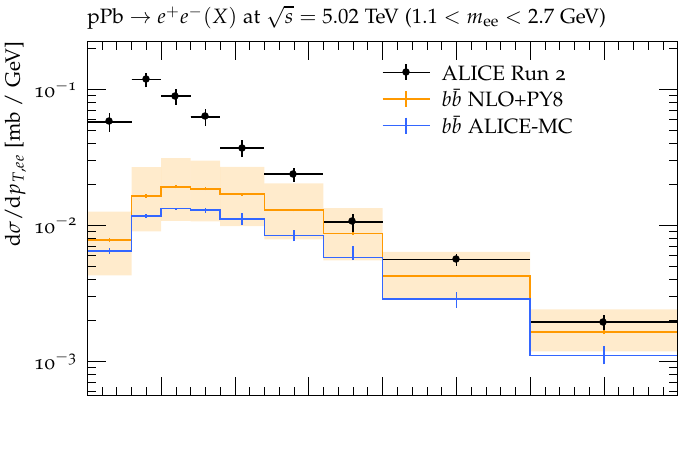}}
        \vspace{-2.30em}

        \includegraphics[width=0.85\textwidth]{{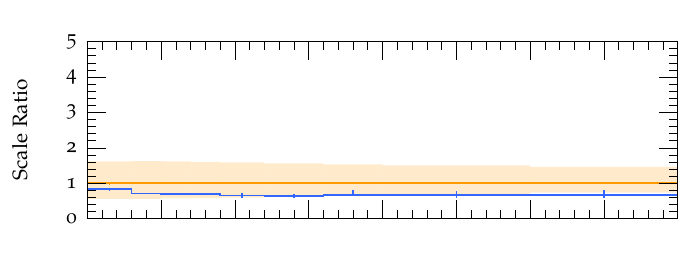}}
        \vspace{-2.30em}

        \includegraphics[width=0.85\textwidth]{{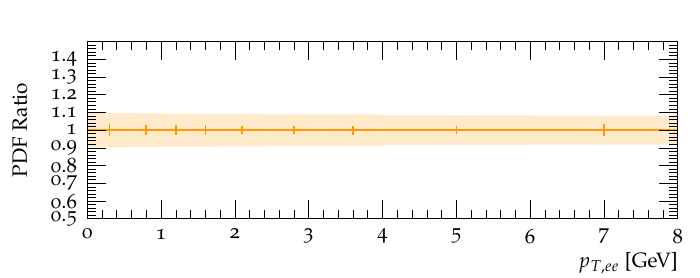}}
        \vspace{-2.30em}
    \end{subfigure}
    \vspace{3em}
    \caption{
        Di-electron transverse momentum spectra in open bottom production $\text{pp, pPb} \to b \bar b(X)$ obtained with  \texttt{POWHEG+Py8} and MSHT20nlo and nCTEQ15HQ PDFs for p and Pb. 
        The second row shows the scale and the third row the PDF uncertainties. 
    }
    \label{fig:bb_imr}
\end{figure}

\begin{figure}
    \centering
    \begin{subfigure}[b]{0.48\textwidth}
        \centering
        \includegraphics[width=0.85\textwidth]{{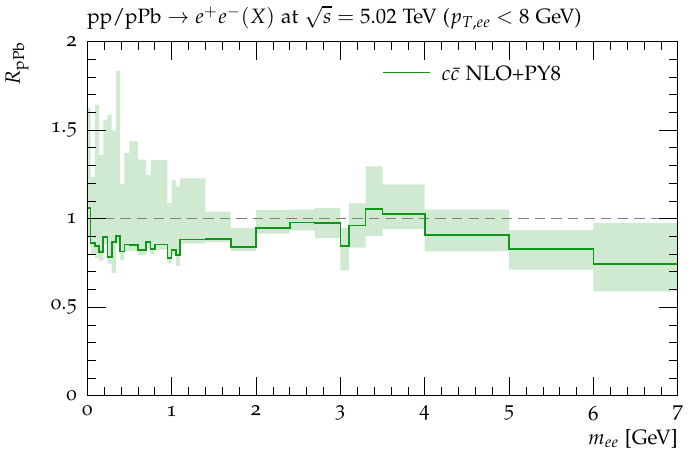}}
        \vspace{-2.30em}
    \end{subfigure}
    \hfill
    \begin{subfigure}[b]{0.48\textwidth}
        \centering
        \includegraphics[width=0.85\textwidth]{{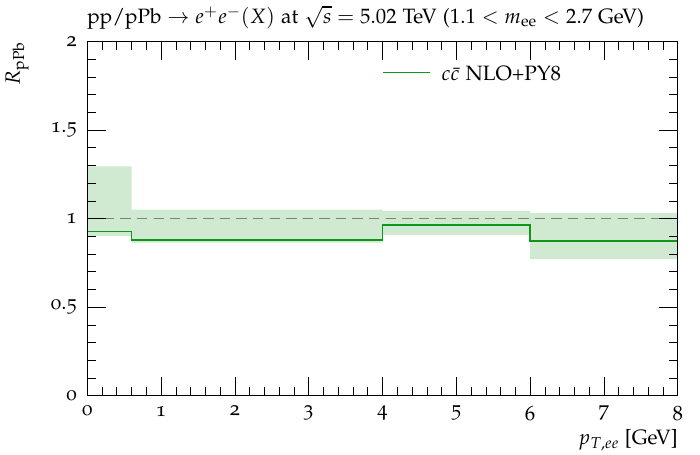}}
        \vspace{-2.30em}
    \end{subfigure}
    \vspace{3em}
    \caption{
        Di-electron pair ratios $R_\text{pPb}$ of the invariant mass (left) and transverse momentum distributions (right) for the $c\bar c$ process.
        The uncertainty bands are correlated scale uncertainties.
        \label{fig:cc_NLO_rpPb}
    }
\end{figure}

\begin{figure}
    \centering
    \begin{subfigure}[b]{0.48\textwidth}
        \centering
        \includegraphics[width=0.85\textwidth]{{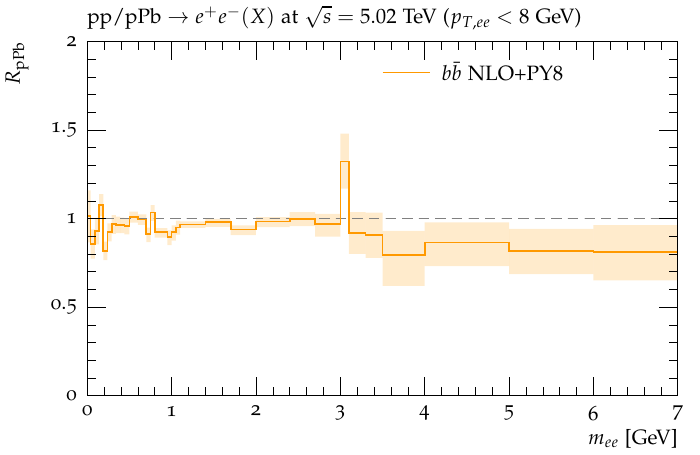}}
        \vspace{-2.30em}
    \end{subfigure}
    \hfill
    \begin{subfigure}[b]{0.48\textwidth}
        \centering
        \includegraphics[width=0.85\textwidth]{{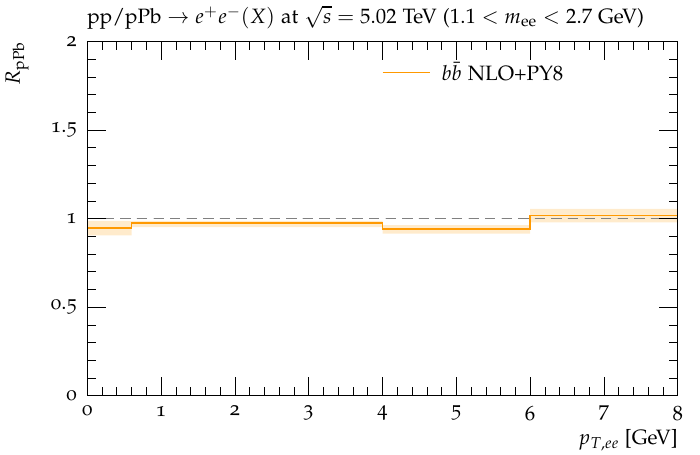}}
        \vspace{-2.30em}
    \end{subfigure}
    \vspace{3em}
    \caption{
        Di-electron pair ratios $R_\text{pPb}$ of the invariant mass (left) and transverse momentum distributions (right) for the $b\bar b$ process.
        The uncertainty bands are correlated scale uncertainties.
        \label{fig:bb_NLO_rpPb}
    }
\end{figure}

Next, we present our predictions for open charm and bottom production, again in pp and pPb collisions at $\sqrt{s} = \SI{5.02}{TeV}$.
The first rows of \Cref{fig:cc_inv,fig:cc_imr} respectively show the absolute predictions for the invariant mass and the transverse momentum spectra of di-electrons from open charm at NLO+PS in green.
In the second rows we show the perturbative scale variations and generally observe very large uncertainties, presumably due to large logarithms related to the charm quark mass, which can exceed a factor of three regardless of the collision system. 
For di-electrons from open bottom production the situation is under much better control and the scale uncertainties 
drop under ${}^{+\SI{60}{\%}}_{-\SI{40}{\%}}$, see \Cref{fig:bb_inv,fig:bb_imr}. 
The third rows in \Cref{fig:cc_inv,fig:cc_imr,fig:bb_inv,fig:bb_imr} show that the PDF uncertainty is small relative to the scale band and under $\pm$\SI{20}{\%} ($\pm$\SI{10}{\%}) for the open charm (bottom) channel for both pp and pPb collisions throughout.
Comparing the pp and pPb PDF uncertainties they increase slightly from pp to pPb in open bottom but not in the open charm production. 
This suggests that very low-$x$ gluons in the nuclear PDF fit (nCTEQ15HQ) have a smaller uncertainty than in the proton PDF fit (MSHT20nlo).  
A realistic PDF uncertainty estimate in pPb collisions should of course include both the proton as well as the nuclear uncertainty dimension, which is however not available in the nCTEQ15HQ fit.

We also compare our predictions to the \texttt{POWHEG+Pythia6} (ALICE-MC) calculations, in blue, and data from ALICE \cite{ALICE:2020mfy}, in black. 
The heavy flavour production is subleading in the LMR, but in the region in $m_{ee} \in [1,2]$ GeV the open charm production almost saturates the data. 
The two predictions generally agree well. There are differences both in normalisation and shape, but they agree within the scale uncertainty band of our prediction for both the open charm and open bottom channel, except for the tails of $m_{ee}$. 
The remaining differences in the shapes of the invariant mass distributions can be explained by the different hadronisation model in \texttt{Pythia} version 6 vs.~version 8 and the use of different, in the case of our predictions more modern, PDF sets.

In the measurements of the di-electron spectra, the total $c\bar{c}$ and $b\bar{b}$ production cross section is determined from the data. The normalisation component of the uncertainty must thus not necessarily be taken into account.
If we were to fix the normalisation of each of the predictions in the scale or PDF variation bands, we would expect the uncertainty due to normalisation to drop out and the uncertainty bands to shrink. 
To what extent this happens is investigated in the next section.

In \Cref{fig:cc_NLO_rpPb,fig:bb_NLO_rpPb} we show the ratio of pPb and pp predictions for charm and bottom production respectively. In the case of open charm the ratios as functions of both $m_{ee}$ and $p_{T,ee}$ are up to fluctuations consistent with one within the scale uncertainty which is significantly reduced in the ratio, a few tens of percent, as compared to the absolute pp and pPb predictions, up to factor three. For open bottom production, the scale uncertainties vanish almost completely and the $R_{pPb}$ ratios are up to fluctuations within 10\% of unity, except in the high tail of $m_{ee}$ where it reaches the value of 0.8.

We find that \texttt{Pythia8} mainly hadronises the charm quarks to D$^\pm$ (\SI{50}{\%}), D$_0$ (\SI{39}{\%}) and D$_s^\pm$ (\SI{7}{\%}) hadrons,
whereas the immediate parent particles of the electrons in open bottom production are mainly D$^\pm$ (\SI{23}{\%}), D$^0$ (\SI{23}{\%}), B$_0$ (\SI{22}{\%}), B$^\pm$ (\SI{20}{\%}), D$^\pm_s$ (\SI{7}{\%}) and B$_s^\pm$ (\SI{4}{\%}) hadrons.

\FloatBarrier

\subsection{Combined analysis of the signal and background\label{sec:combination}}

\begin{figure}
    \centering
    \begin{subfigure}[b]{0.49\textwidth}
        \centering
        \includegraphics[width=0.85\textwidth]{{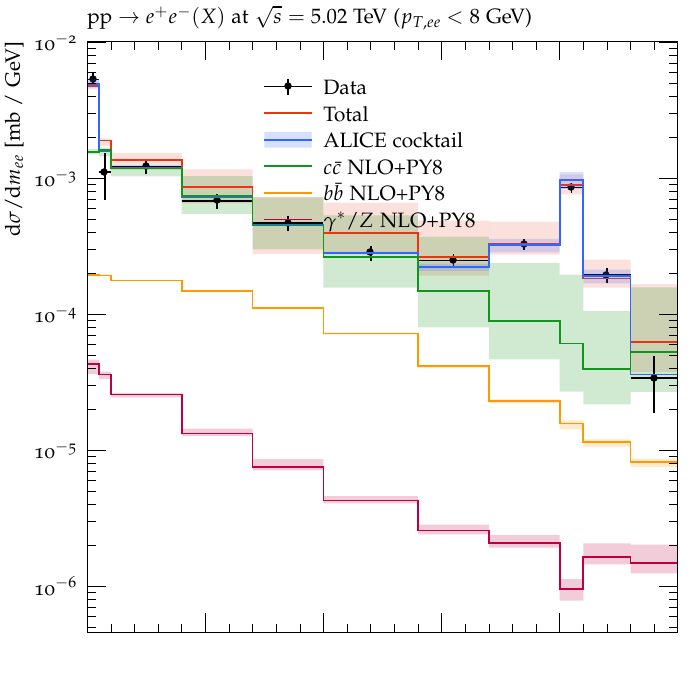}}
        \vspace{-2.30em}

        \includegraphics[width=0.85\textwidth]{{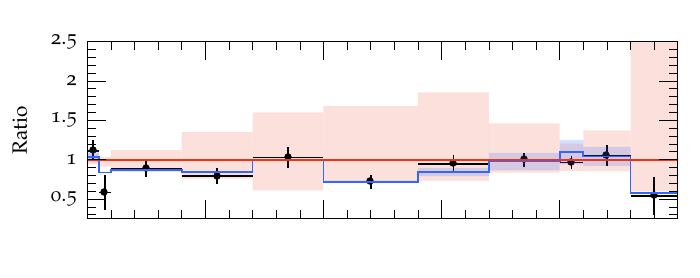}}
        \vspace{-2.30em}

        \includegraphics[width=0.85\textwidth]{{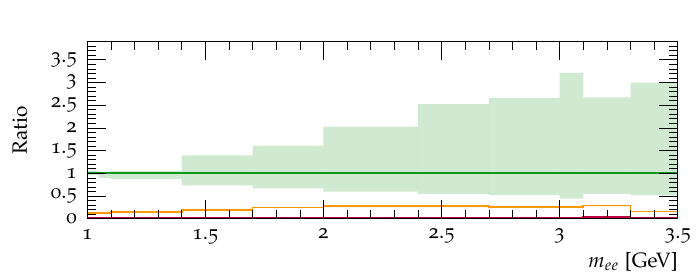}}
        \vspace{-2.30em}
    \end{subfigure}
    \hfill
    \begin{subfigure}[b]{0.49\textwidth}
        \centering
        \includegraphics[width=0.85\textwidth]{{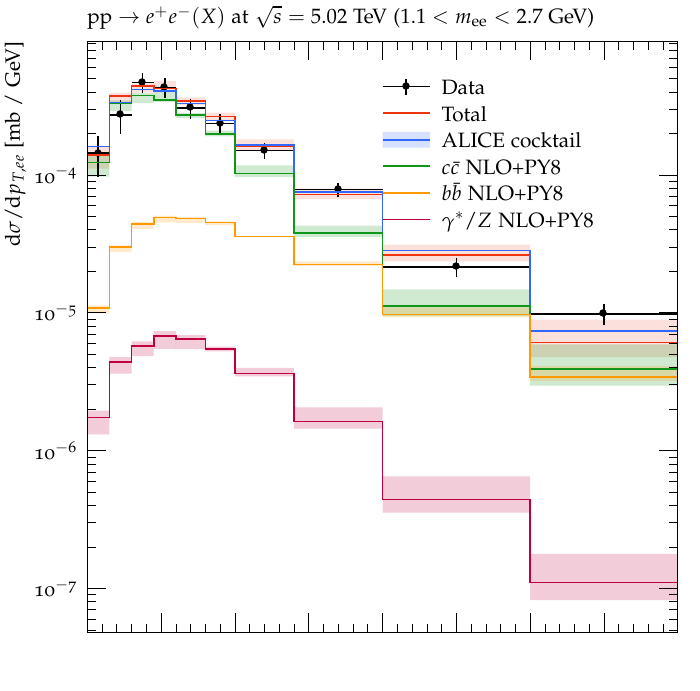}}
        \vspace{-2.30em}

        \includegraphics[width=0.85\textwidth]{{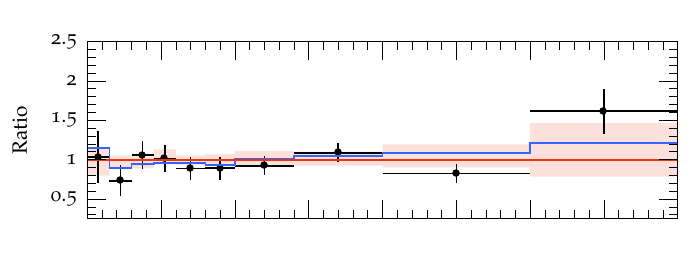}}
        \vspace{-2.30em}

        \includegraphics[width=0.85\textwidth]{{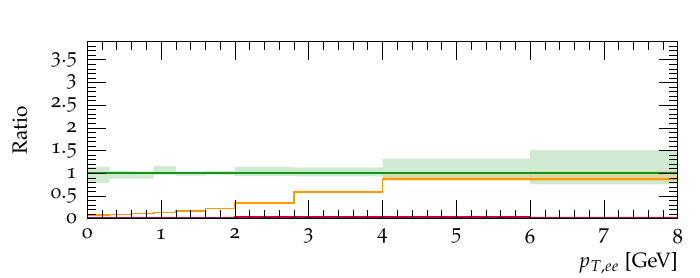}}
        \vspace{-2.30em}
    \end{subfigure}
    \vspace{3em}
    \caption{
        Comparison of the cocktail used by ALICE against our new cocktail including the Drell-Yan process with an extra jet.
        The shaded band is the scale shape uncertainty.
        The blue shaded band on the cocktails is the uncertainty given by ALICE collaboration on their cocktail.
        The red shaded band combines scale shape and ALICE uncertainties.
    }
    \label{fig:total}
\end{figure}

In this section we consider our new prediction for the virtual photon and the revised predictions for open heavy-flavour production together. 
The first rows on both panels in \Cref{fig:total} show our absolute predictions for virtual photon production in purple, $b \bar b$ in yellow and $c \bar c$ in green, whereas the third rows show their ratio to the largest channel ($c \bar c$).
The normalisation of the heavy flavour continuum is typically determined via double differential fits in $m_{ee}$ and $p_{T,ee}$ in the intermediate mass regime \cite{ALICE:2020mfy} or via the measured and extrapolated total cross section \cite{ALICE:2018gev,Averbeck:2011ga}.
In this section, thus, we also normalise our predictions to the same cross sections that ALICE does, \cite{ALICE:2020mfy}, 
by multiplying by the two overall normalisations factors, $f_{c\bar c}$ and $f_{b\bar b}$, for open charm and open bottom production respectively. 
We determine those factors to be $f_{c\bar c} = 2.5$ and $f_{b\bar b} = 0.75$.
Similarly, we normalise the virtual photon contribution by a factor $f_{\gamma/Z}=0.06$. This value will be explained in what follows.
Such a normalisation procedure also impacts our predictions for uncertainties. 
In fact, one needs to normalise each of our predictions in the seven-point scale or PDF variations individually.
This means that the uncertainty due to normalisation cancels out and only that due to change of shape remains.
We observed that the different error sets of the PDF do not give a different shape and thus nearly vanish when normalised such that PDF uncertainties can be neglected.
The $c \bar c$ contribution dominates across the whole $m_{ee}$ and $p_{T,ee}$ range, with $b \bar b$ coming in second reaching up to 30\% in the bin at 2 GeV in the invariant mass and up to 100\% in the high tail of the transverse momentum spectrum. 
The virtual photon contributes under 10\% across the whole range.
The $c\bar c$ production yields an appreciably softer transverse momentum spectrum, dominating the low $p_T$ region, as compared to the $b \bar b$ which starts to contribute significantly at around 2 GeV and eventually matches the $c\bar c$ channel. 
The shapes of the invariant mass spectra for the two heavy flavour production modes, instead, are relatively similar, except below $m_{ee} =1.5$ GeV, where the bottom spectrum flattens out whereas the charm does not. 
Both the invariant mass and the transverse momentum spectra of the virtual photon contribution are relatively flat, as compared to those of open charm production.
Thanks to our normalisation procedure, the uncertainty bands are now considerably reduced, compared to \Cref{fig:cc_inv,fig:cc_imr,fig:bb_inv,fig:bb_imr}. 
However, the remaining {\em shape} uncertainty, dominated by the scale variation, is still considerable especially in the $c\bar c$ channel where it goes from about $\pm$10\% at $m_{ee} = m_c$ to about ${}^{+150\%}_{-50\%}$ in the high tail. 
This shape uncertainty is in fact significantly larger than the statistical and systematic uncertainty as well as the uncertainty on the branching ratios into electrons for $ c \bar c$ production, respectively, which were found to be 11\%, 5\% and 22\% in a previous study by ALICE. 

We now want to combine our predictions for virtual photon and open heavy flavour production into a new prediction for the hadronic cocktail. 
To this end we start by isolating the contributions from light flavour (LF) and $J/\psi$ from ALICE cocktail from Ref.\,\cite{ALICE:2020mfy}, simply by subtracting the $c\bar{c}$ and $b\bar{b}$ ALICE-MC predictions bin-by-bin. 
We then sum all of the contributions together, optionally including virtual photons, with a normalisation for each contribution left as a free parameter, and perform several fits to all publicly available ALICE data in which we extract the values of $f_{c\bar{c}}$, $f_{b\bar{b}}$, $f_{\text{LF}+J/\psi}$ and $f_{\gamma/Z}$. 
Our fit results are summarised in \Cref{tab:total}.
In the first column we report the results of the original ALICE fit, using the ALICE-MC predictions. The fit determines $f_{c\bar{c}} = 2.5$ and $f_{b\bar{b}}=0.75$ and describes the data well with $\chi^2/\text{ndf} = 0.9$.
We repeat this fit twice, first with all the data points in the LMR and in the IMR (i.e.~one invariant mass spectrum in its full range and two transverse momentum spectra for each pp and pPb collisions) we had access to (npts = 89 as compared to npts = 123 in the ALICE fit) and only our predictions for the background, and then the second time only in the IMR (npts = 53, i.e.~one invariant mass spectrum in a reduced range and one transverse momentum spectrum for each pp and pPb collisions) but now also with our prediction for the virtual photons.
Each time we offer two alternatives, one in which the shape uncertainties are not considered (``central'', $\chi^2 = \sum_i (x_i-m_i)^2/\Delta m_i^2$) and one in which they are considered in a fully uncorrelated manner (``shape'', $\chi^2 = \sum_i (x_i-m_i)^2/(\Delta m_i^2 + \Delta x_i^2)$).  
The uncertainties on the normalisation coefficients $f$, reported in round brackets in columns 2 to 5 of \Cref{tab:total}, are the Hesse errors obtained from the $\chi^2$ minimisation by \texttt{iminuit}~\cite{MINUIT,IMINUIT}. 
The uncertainty on $\sigma_{c\bar c}$ and $\sigma_{b\bar b}$ in the first column, also reported in round brackets, are instead obtained by combining all uncertainties from Table 3 of Ref.~\cite{ALICE:2020mfy} in quadrature. 
Correspondingly, the uncertainties on $\sigma_{c\bar c}$ and $\sigma_{b\bar b}$ in columns 2 to 5 are obtained by rescaling the uncertainty in column 1 by the ratio of the newly extracted cross section to that of column 1.
In our first fit in columns 2 and 3, we get a very good data description with $\chi^2/\text{ndf}$ equal to 0.95 and 0.74, respectively. 
The normalisation coefficient $f_{c\bar{c}}$ increases which in turn leads to an increase of extracted values of the $c\bar{c}$ cross sections from 756 $\mu$b $\to$ 788 $\mu$b and to 888 $\mu$b, respectively.
The normalisation coefficient $f_{b\bar{b}}$ is instead slightly decreased. 
The relative fit uncertainties on the normalisation coefficients are small ($\sim$ 5\%) as compared to the uncertainty on the total extracted cross section reported by ALICE ($\sim$ 25\%) for the $c\bar{c}$ channel, but not for $b\bar{b}$ ($\sim$ 18-25\% vs. $\sim$ 14\%).
In our second fit in columns 4 and 5, the addition of the virtual photon contribution pulls the fit back in the direction of the original fit in column 1 reducing the factor $f_{c\bar c}$ and increasing $f_{b \bar b}$.
The fit prefers a smaller value of the virtual photon normalisation of 23\% and 6\% that is however still compatible with 1.0 within $3\sigma$, in particular in view of the fact that in this kinematic region with very small values of photon virtuality and transverse momentum, the dependence of the cross section on the choice of the factorisation and renormalisation scales is significant.
Thus the sensitivity to the virtual photon contribution is limited and strongly dependent on the uncertainties of the background contributions.
All our fits prefer a reduced fraction of the light flavour and $J/\psi$ contributions, more notably in fits that include the virtual photon contribution. 
Ultimately, however, the fact whether the {\em shape} scale uncertainty is included in the fit or not plays a more important role. 
All in all, the light flavour and $J/\psi$ contribution is compatible with 1.0 within 2 standard deviations and in agreement with previous studies.
Owing to the large branching ratio uncertainty the extracted charm and bottom cross sections are also compatible within one sigma across all five fits.
In general, the inclusion of shape uncertainties increases these uncertainties for every fitting parameter, but most significantly, as expected, for $f_{c\bar c}$.
The best description of the data, i.e. the smallest $\chi^2/\text{ndf}$, is obtained in the fit which takes both the virtual photon and the shape uncertainties into account.

In \Cref{fig:total}, we compare our best prediction for the hadronic cocktail (corresponding to fit 2) denoted {\em Total} and shown in red, to the hadronic cocktail from ALICE, in blue, and to data in black. 
Our prediction for the hadronic cocktail agrees with the previous prediction from ALICE and describes the measurement well. 
Our central fit describes the data less well than the central fit from ALICE, deviating from the data by up to 30\%, however only one point is more than 1$\sigma$ away, not considering the theoretical uncertainties. 
However, our hadronic cocktail deviates from the one by ALICE considerably when it comes to uncertainties. 
Whereas ALICE's uncertainties are well below $\pm$5\% in the low tail increasing to about $\pm$10\% in the high tail, our uncertainties reach up to $\pm$50\% and almost never drop under $\pm$10\%. 
This striking difference is due to the perturbative scale shape uncertainties in the $c\bar c$ production, which were not considered in the ALICE analysis.
The inclusion of the virtual photon contribution does not change the shape of the total result appreciably relative to the data uncertainties.

\begin{table}
    \centering
    \caption{
        Different normalisation parameters $f$, $\chi^2/\text{ndf}$ and heavy-quark cross sections for the different combinations of processes contributing to di-electron production obtained from a fit with \texttt{iminuit} \cite{MINUIT,IMINUIT}.
    }
    \label{tab:total}
    \begin{tabular}{cccccc}
        \hline
        \textbf{} & ALICE cocktail& \multicolumn{2}{c}{LF+$J/\psi$+$b\bar b$+$c\bar c$} & \multicolumn{2}{c}{LF+$J/\psi$+$b\bar b$+$c\bar c$+$\gamma/Z$}  \\
        dataset & pp \& pPb & \multicolumn{2}{c}{pp \& pPb} & \multicolumn{2}{c}{pp \& pPb (no LMR)}\\
        uncertainty & central & central & shape & central & shape \\
        \hline
        $f_{c\bar c}$                          & 2.50               & \SI{2.61(16)}{}          & \SI{2.93(19)}{}  & \SI{2.35(24)}{}& \SI{2.86(31)}{}            \\
        $f_{b\bar b}$                          & 0.75               & \SI{0.60(11)}{}          & \SI{0.48(12)}{}  & \SI{0.63(13)}{}& \SI{0.57(15)}{}            \\
        $f_{\text{LF}+J/\psi}$                 & 1.0                & \SI{0.96(3)}{}           & \SI{0.93(3)}{}   & \SI{0.91(6)}{} & \SI{0.88(8)}{}            \\
        $f_{\gamma/Z}$                         & -                  & -                        & -                & \SI{0.23(17)}{}& \SI{0.06(20)}{}            \\
        $\chi^2/\text{ndf}$                    & $110.9/123$        & $84.2 / 89$              & $65.9 / 89$      & $47.7 / 49$    & $33.3 / 49$     \\
                                               & $0.90$             & $0.95$                   & $0.74$           & $0.97$         & $0.68$          \\
        $\dd{\sigma}_{c \bar c}/\dd{y}|_{y=0}$ [\si{\mu\barn}]&\SI{756(188)}{}&\SI{788(196)}{}& \SI{888(221)}{}&\SI{709(176)}{}    & \SI{866(215)}{} \\
        $\dd{\sigma}_{b \bar b}/\dd{y}|_{y=0}$ [\si{\mu\barn}]&\SI{28(5)}{}   &\SI{22(3)}{}   & \SI{22(3)}{}    &\SI{24(3)}{}       & \SI{21(3)}{}    \\
        \hline
    \end{tabular}
\end{table}

\begin{figure}
    \centering
    \begin{subfigure}[b]{0.49\textwidth}
        \centering
        \includegraphics[width=0.85\textwidth]{{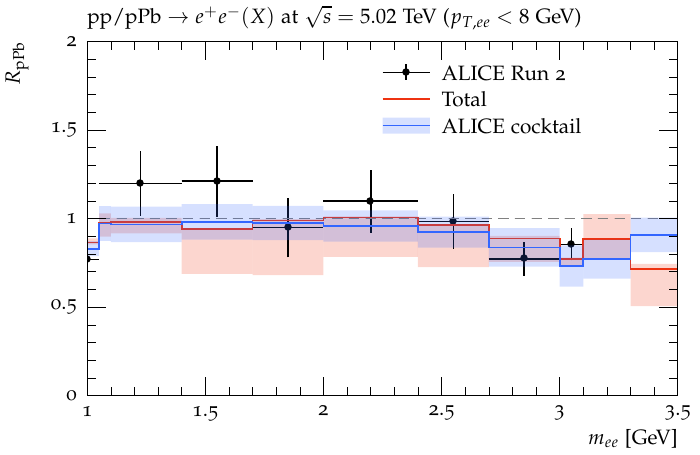}}
        \vspace{-2.30em}
    \end{subfigure}
    \hfill
    \begin{subfigure}[b]{0.49\textwidth}
        \centering
        \includegraphics[width=0.85\textwidth]{{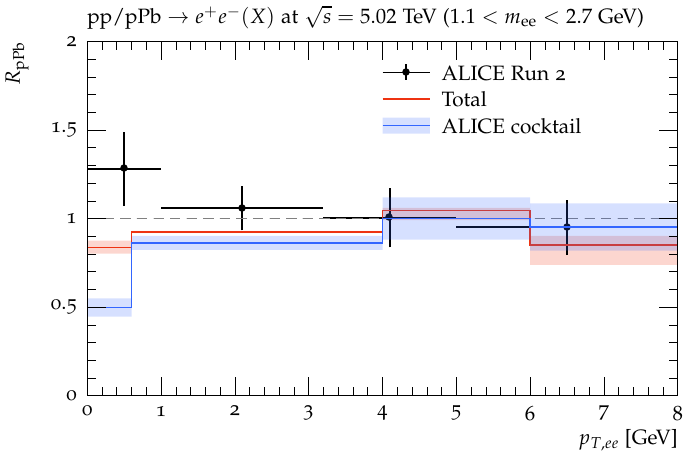}}
        \vspace{-2.30em}
    \end{subfigure}
    \vspace{3em}
    \caption{
        $R_\text{pPb}$ in invariant mass (left) and transverse momentum (right) for the total cocktail.
        For the Total curve the uncertainty band are correlated scale uncertainties of the $c\bar c$,$b\bar b$ and $Zj$ process.
        The uncertainty band of the ALICE cocktail is computed solely from the lead uncertainty given by ALICE in the nominator of the ratio.
    }
    \label{fig:total_rpPb}
\end{figure}

Finally, in \Cref{fig:total_rpPb}, we show the ratios our pPb and pp compared to those from ALICE and the data. Both ratios were obtained by simply dividing the pPb prediction by the pp one after appropriate rebinning. The $R_{pPb}$ predictions agree very well across the whole range of $m_{ee}$ and $p_{T,ee}$ spectra except in the first transverse momentum bin where our prediction is closer to the data. In the intermediate range of the invariant mass spectrum, our prediction features appreciably larger uncertainties as a result of the open charm production scale shape uncertainties not completely canceling in the ratio. 

\FloatBarrier

\subsection{Initial- and final-state nuclear effects in PbPb collisions \label{sec:leadLead}}

\begin{figure}
    \centering
    \begin{subfigure}[b]{0.48\textwidth}
        \includegraphics[width=\textwidth]{{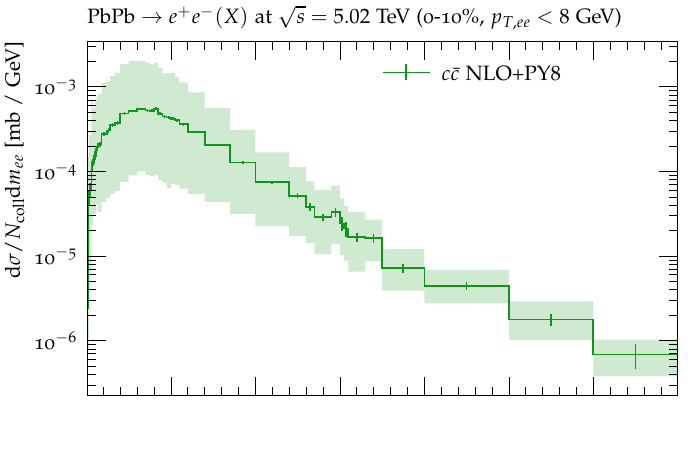}}
        \vspace{-4em}

        \includegraphics[width=\textwidth]{{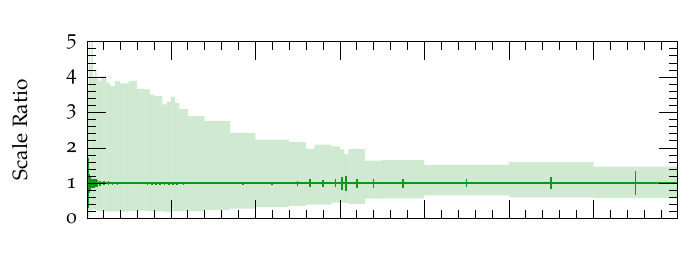}}
        \vspace{-4em}

        \includegraphics[width=\textwidth]{{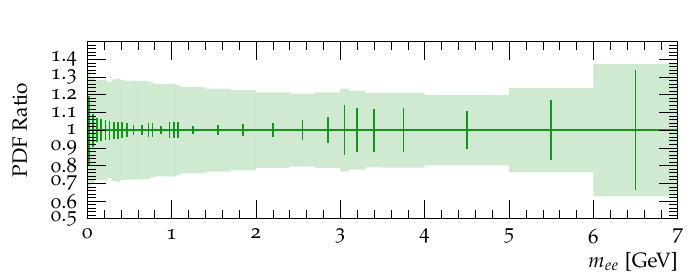}}
    \end{subfigure}
    \begin{subfigure}[b]{0.48\textwidth}
        \includegraphics[width=\textwidth]{{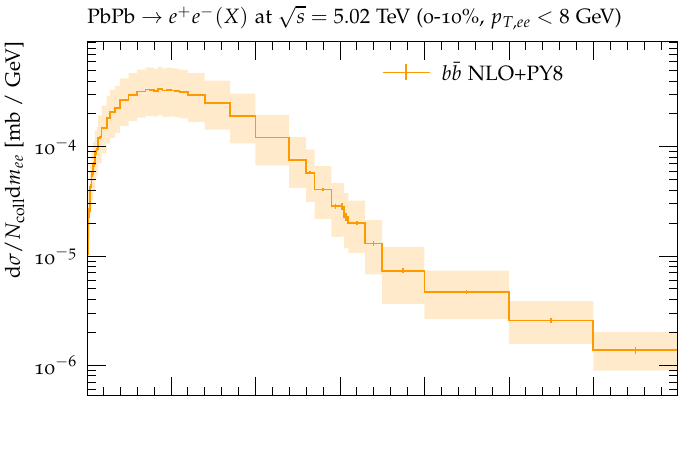}}
        \vspace{-4em}

        \includegraphics[width=\textwidth]{{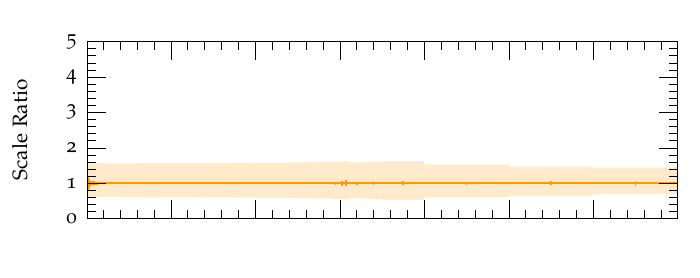}}
        \vspace{-4em}

        \includegraphics[width=\textwidth]{{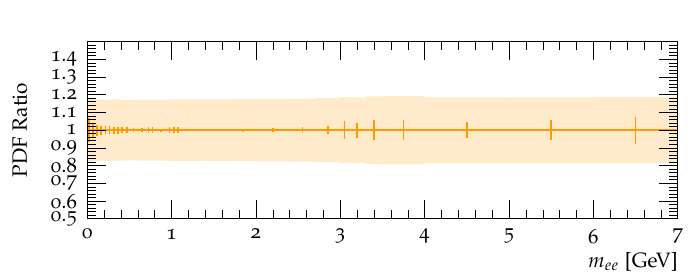}}
    \end{subfigure}
    \caption{
        Invariant mass spectrum to open charm production $\text{PbPb} \to c \bar c(X), b \bar b(X)$ using \texttt{POWHEG} and \texttt{Pythia} with nCTEQ15HQ as PDFs. 
        The second row is scale uncertainty and the third row is PDF uncertainty.
    }
    \label{fig:PbPb_inv}
\end{figure}

\begin{figure}
    \centering
    \begin{subfigure}[b]{0.48\textwidth}
        \includegraphics[width=\textwidth]{{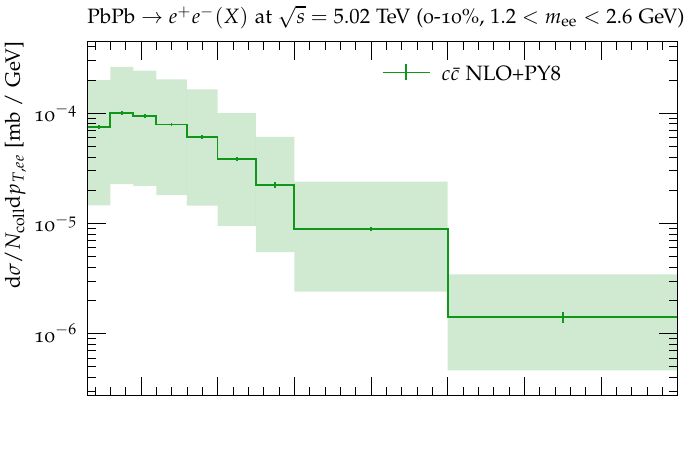}}
        \vspace{-4em}

        \includegraphics[width=\textwidth]{{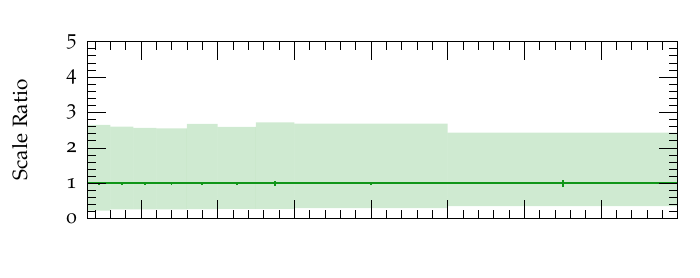}}
        \vspace{-4em}

        \includegraphics[width=\textwidth]{{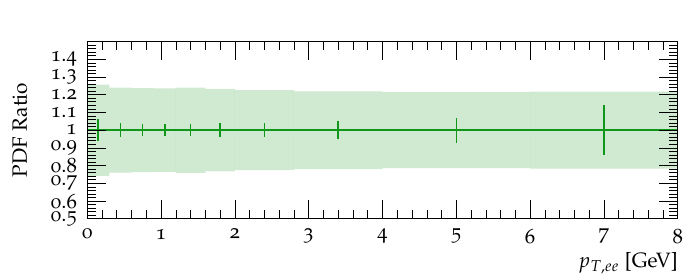}}
    \end{subfigure}
    \begin{subfigure}[b]{0.48\textwidth}
        \includegraphics[width=\textwidth]{{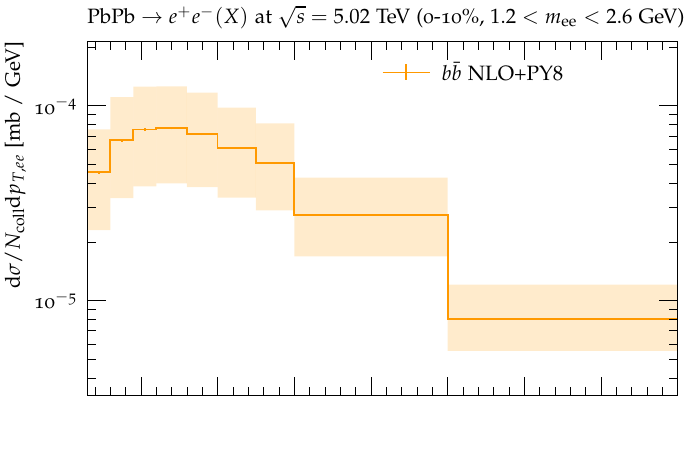}}
        \vspace{-4em}

        \includegraphics[width=\textwidth]{{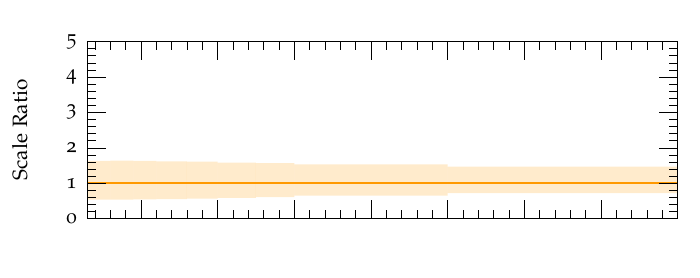}}
        \vspace{-4em}

        \includegraphics[width=\textwidth]{{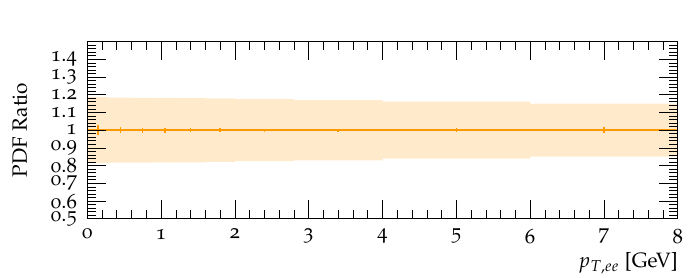}}
    \end{subfigure}
    \caption{
        Transverse momentum spectrum in the intermediate invariant mass region to open charm production $\text{PbPb} \to c \bar c(X), b \bar b (X)$ using \texttt{POWHEG} and \texttt{Pythia} with nCTEQ15HQ as PDFs.
        The second row is scale uncertainty and the third row is PDF uncertainty.
    }
    \label{fig:PbPb_imr}
\end{figure}

In this section we focus on PbPb collisions at 5.02 TeV. 
Our predictions for the invariant mass and transverse momentum spectra per collision of open heavy flavour production are presented in \Cref{fig:PbPb_inv,fig:PbPb_imr}, respectively. The predictions are qualitatively very similar to the pp and pPb cases with very large scale variation uncertainties of up to a factor four in the case of $c\bar{c}$ and ${}^{+60\%}_{-40\%}$ for $b\bar{b}$. The PDF uncertainties, as expected, are now instead larger reaching up to $\pm30\%$ and $\pm20\%$ for $c\bar{c}$ and $b\bar{b}$ respectively. 
Even though a measurement of both spectra in PbPb collisions is now available in Ref.~\cite{ALICE:2023jef}, we weren't able to compare our predictions to data since it is not yet available in {\tt HEPdata}. 

\begin{figure}
    \centering
    \begin{subfigure}[b]{0.32\textwidth}
        \includegraphics[width=\textwidth]{{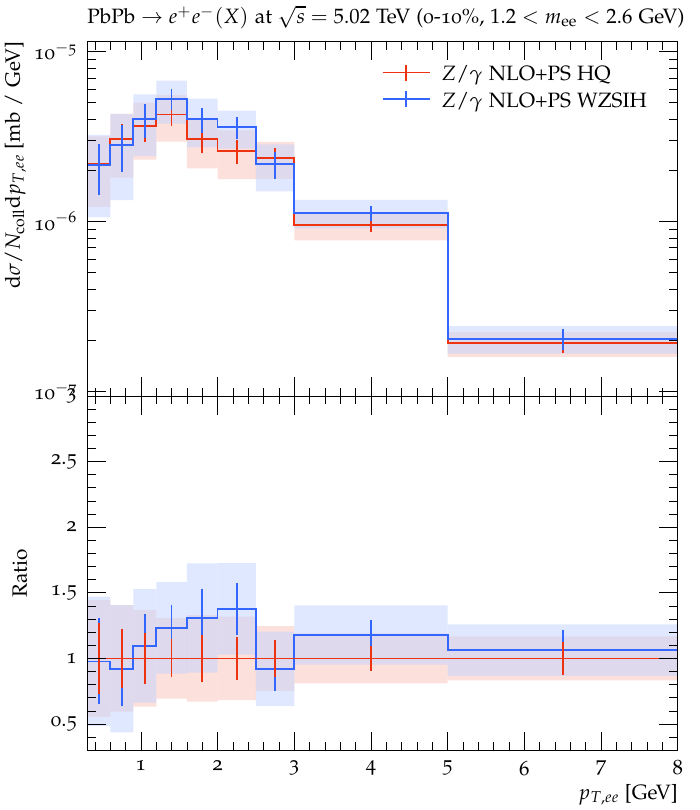}}
    \end{subfigure}
    \hfill
    \begin{subfigure}[b]{0.32\textwidth}
        \includegraphics[width=\textwidth]{{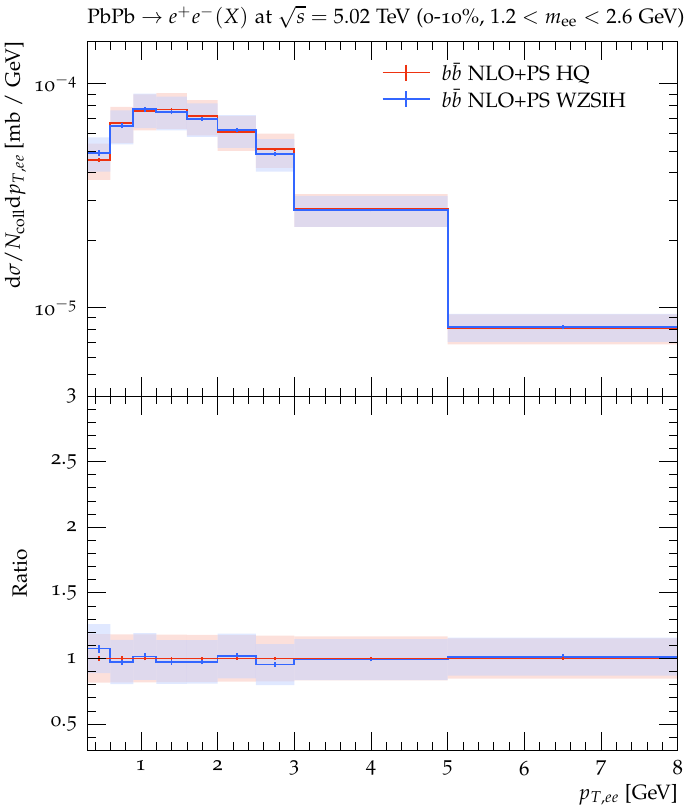}}
    \end{subfigure}
    \hfill
    \begin{subfigure}[b]{0.32\textwidth}
        \includegraphics[width=\textwidth]{{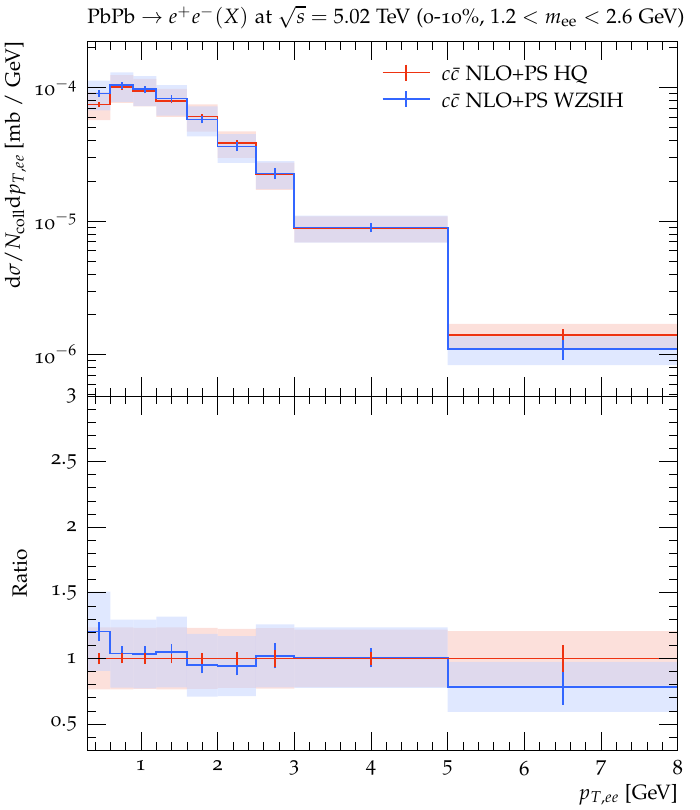}}
    \end{subfigure}
    \caption{
        Comparison of PbPb nuclear PDF uncertainties between nCTEQs WZSIH and HQ PDF sets.
        Left: Drell-Yan with one jet.
        Middle: $c \bar c$-production. 
        Right: $b \bar b$-production.
        The shaded band is the PDF uncertainty.
        The vertical bars are the statistical uncertainty of the simulation.
    }
    \label{fig:ncteq_hq}
\end{figure}

The nCTEQ15HQ \cite{Duwentaster:2022kpv} nPDFs improve on the previous nCTEQ15WZSIH \cite{Duwentaster:2021ioo} fit by including the heavy quark (HQ) data, both the open HQ hadrons and  quarkonia. 
It is thus very interesting to see what impact the inclusion of heavy quark data in the PDF has on virtual photon and heavy quark production in the kinematic regime considered here.
\Cref{fig:ncteq_hq} shows the $p_{T,ee}$ spectra of the production of virtual photons on the left,  $c \bar c$ in the middle and  $b \bar b$ on the right in simulated \SI{5.02}{TeV} PbPb collisions.
The inclusion of the HQ data reduces the virtual photon production cross section by about 20\% and hardens the spectrum of open charm production gently but leaves the open bottom spectrum mostly unaffected. 
Because the uncertainties are large, however, all the predictions are still consistent within their uncertainties. 
Curiously though, we note that the inclusion of the HQ data reduces the uncertainties in the virtual photon and in the open heavy flavour production predictions only marginally.
This is counterintuitive.
We would expect the open heavy flavour dominated by gluon initiated channels to profit from the reduction of uncertainties in the gluon distribution of the nCTEQ15HQ fit in low-$x$ as compared to nCTEQ15WZSIH and the mostly quark initiated virtual photon production to not change. 
In order to understand this one has to also consider the momentum fractions relevant for our kinematic regime: the heavy quark process receives most of its contribution at $x\sim10^{-3}$ and the virtual photon production at $x\sim10^{-4}$, where sea quarks that are correlated with the gluon dominate. 

\begin{figure}
    \centering
    \begin{subfigure}[b]{0.3\textwidth}
        \includegraphics[width=5cm]{{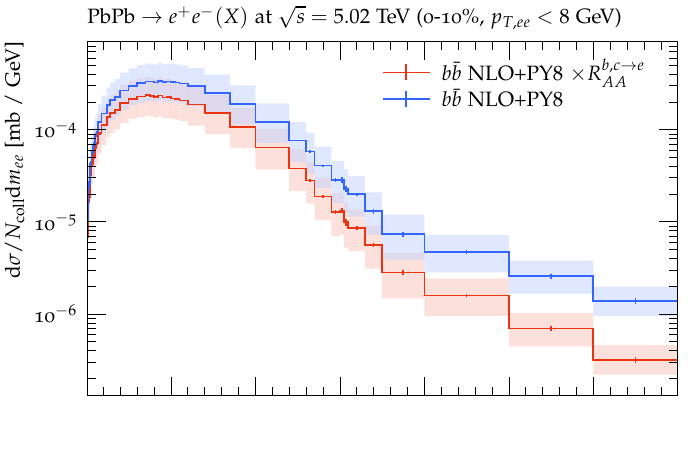}}
        \vspace{-3.25em}

        \includegraphics[width=5cm]{{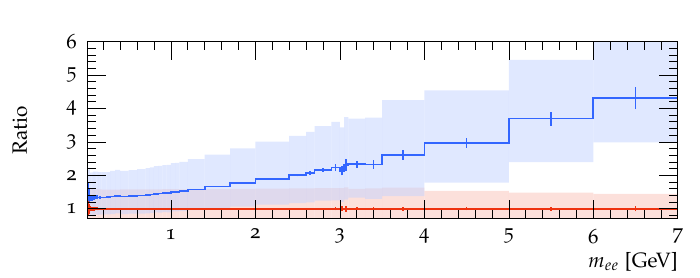}}
    \end{subfigure}
    \quad\quad\quad
    \begin{subfigure}[b]{0.3\textwidth}
        \includegraphics[width=5cm]{{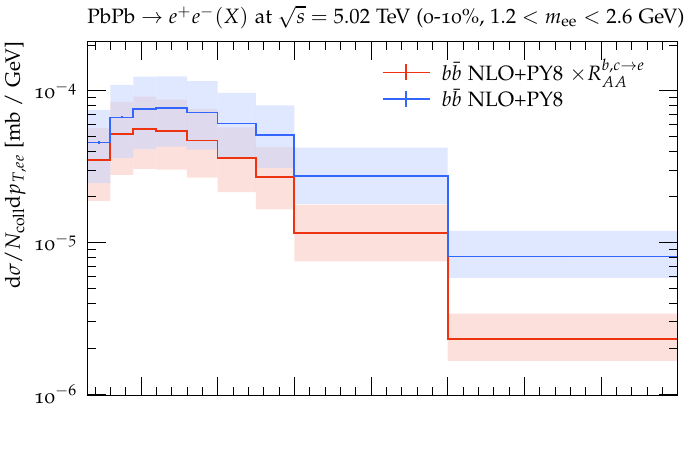}}
        \vspace{-3.25em}

        \includegraphics[width=5cm]{{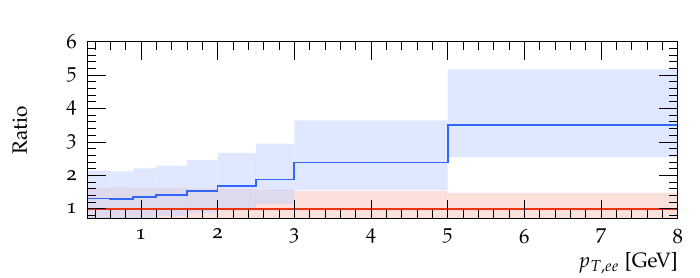}}
    \end{subfigure}
    \caption{ 
        Open bottom production $\text{PbPb} \to b\bar b(X)$ with and without including a $R^{b,c\to e}_{AA}$ factor. 
        The lead nuclear PDF in use is nCTEQ15HQ.
    }
    \label{fig:RAA}
\end{figure}

There is another important correction to the processes $\text{PbPb}\to c\bar c(X), b\bar b(X)$, namely in Ref.~\cite{ALICE:2019nuy} a suppression of hadronic decays involving transitions $c \to e$ and $b \to e$ was measured relative pp collisions, i.e.~$R_{AA}^{c/b\to e}(p_T)$ \cite{Song:2018xca}.
This correction depends on the centrality and can give a reduction of up to \SI{50}{\%} in collisions of centrality \SIrange{0}{10}{\%} and \SIrange{30}{50}{\%} and the effects disappear in collisions of centrality \SIrange{60}{80}{\%}.
The inclusion of this final state effect is multiplicative and due to the dependence on $p_T$ it doesn't only affect the normalisation but also the shape of the distribution.
Similar effects for $R_{pA}$ are possible, but have not yet been measured with sufficient precision \cite{ALICE:2019nuy}.
\Cref{fig:RAA} shows the effect of including this nuclear modification factor $R^{b,c\to e}_{AA}$ on $b\bar b$ production in \SIrange{0}{10}{\%} centrality collisions.
The correction by a factor of 2 to 3 is well above the largest source of uncertainty so far, which is the scale uncertainty, as we have seen in the PbPb case in \Cref{fig:PbPb_imr}.
The inclusion of such a factor and its uncertainty is therefore necessary to obtain a reliable prediction for the $b\bar b$ production in PbPb collisions, especially if the normalisation is to be determined by a fit.

\FloatBarrier

\section{Conclusion}
\label{sec:concl}

In this paper we presented a new calculation of the production of virtual photons in pp, pPb and PbPb collisions and compared it with ALICE measurements.
We used \texttt{POWHEG\,BOX\,V2} to calculate the Drell-Yan process in association with at least one jet through to NLO QCD and matched it to {\tt Pythia8}.
The produced photon converts into an electron-positron pair, which is then detected by the ALICE detector.
This signature is in the invariant mass range of \SIrange{1}{3}{GeV} much more likely to be produced by the correlated semileptonic decay of open heavy charm and bottom hadrons.
Therefore, we recalculated predictions for those too with more recent PDFs and nuclear PDFs, and performed a thorough study of the theoretical uncertainties.
As the spectra's normalisations are determined from a fit to the experimental data we introduced {\em shape uncertainties} in which the large scale uncertainties associated with this kinematic regime are significantly reduced and the PDF uncertainties become negligible.
By repeating the ALICE analysis and fitting the normalisations of the heavy quark, light flavour and $J/\psi$ contributions, we have found that the light flavour contributions reported by ALICE are compatible with their fixed normalisation of 1.
Unsurprisingly, the remaining large shape scale uncertainties result in relaxed constraints on the normalisation of the heavy quark contributions.
The addition of the virtual photon contribution to the fit results in a small reduction of the normalisation of the other contributions.
The photon normalisation is reduced significantly, but still compatible with 1 within $3\sigma$ as it constitutes the smallest part of the total combined prediction.
With the new virtual photon contribution and a more rigorous treatment of theoretical uncertainties our prediction for the hadronic cocktail represents a more solid basis for ALICE to extract the long sought QGP contribution from thermal photon radiation. 
All our predictions are available in the YODA format as ancillary files.
The cold nuclear matter effects were included throughout via nuclear PDFs and we also showed the impact of the nuclear modification factor $R^{b,c\to e}_{PbPb}$ in PbPb collisions, which is likely to be necessary for comparing theoretical predictions to data.
Moreover we investigated the impact of including heavy quark data in the PDF fit by switching between nCTEQ15HQ and nCTEQ15WZSIH nPDF sets and found the effect to be marginal.

It would be very interesting to explore the possibility of using the MiNLO method for the Drell-Yan+jet process to get a better handle on the scale uncertainties and extend our predictions into the low invariant mass region below 1 GeV where it could eventually be compared to the Kroll-Wada approximation.
We look forward to upcoming ALICE publications on dileptons in pp at $13.6$ TeV and PbPb at $5.36$ TeV.

\acknowledgments
T.J.~and A.P.N.~are grateful to Raphaelle Bailhache and Jerome Jung for insightful discussions. This work has been supported by the BMBF under contract 05P21PMCAA and by the DFG through the Research Training Network 2149 “Strong and Weak Interactions - from Hadrons to Dark Matter”.

\paragraph{Open Access.} 
This article is distributed under the terms of the Creative Commons Attribution License (CC-BY-4.0), which permits any use, distribution and reproduction in any medium, provided the original authors(s) and source are credited.

\bibliography{References}

\end{document}